\documentclass{jfm}

\usepackage{graphicx,
natbib,
upmath,
bm,
amsmath,
amssymb,
amsbsy,
color}
\usepackage[makeroom]{cancel}

\newcommand\NavSto{Navier--Stokes}

\newcommand\Rey{\mbox{\textit{Re}}}
\newcommand\Ret{\Rey_\tau}

\newcommand{\ie}{i.e.\ }

\newcommand{\CC}{\mathrm{c.c.}}
\newcommand\cd{\mathrm{d}}

\newcommand\ce{\mathrm{e}}
\newcommand\ci{\mathrm{i}}

\newcommand\cc{complex-conjugate}
\newcommand\oned{one-di\-men\-sion\-al}
\newcommand\twod{two-di\-men\-sion\-al}

\title[Streamwise-varying steady transpiration control
in turbulent pipe flow]{Streamwise-varying steady transpiration control
in turbulent pipe flow}

\author[F. G\'omez and others]{F. G\'omez$^1$\thanks{Email address for
correspondence: francisco.gomez-carrasco@monash.edu},
  H. M. Blackburn$^{1}$, M. Rudman$^1$,\break A. S. Sharma$^2$
  and B. J. McKeon$^3$}

\affiliation{$^1$Department of Mechanical and Aerospace Engineering,\\ 
Monash University, Victoria 3800, Australia \\[\affilskip]
$^2$Faculty of Engineering and the Environment, \\ 
University of Southampton,
Southampton SO17 1BJ, UK \\[\affilskip]
$^3$Graduate Aerospace Laboratories, \\ 
California Institute of Technology, Pasadena, CA 91125, USA
}

\date{?; revised ?; accepted ?. - To be entered by editorial office}

\begin{document}

\maketitle

\begin{abstract}
The effect of streamwise-varying steady transpiration on turbulent
pipe flow is examined using direct numerical simulation at fixed
friction Reynolds number $\Ret=314$.
The streamwise momentum equation reveals three physical mechanisms
caused by transpiration acting in the flow: modification of Reynolds
shear stress, steady streaming and generation of non-zero mean
streamwise gradients. The influence of these mechanisms has been
examined by means of a parameter sweep involving transpiration
amplitude and wavelength. The observed trends have permitted
identification of wall transpiration configurations able to reduce or
increase the overall flow rate $-36.1$\% and $19.3$\% respectively.
Energetics associated with these modifications are presented.

A novel resolvent formulation has been developed to investigate the
dynamics of pipe flows with a constant cross-section but with
time-mean spatial periodicity induced by changes in boundary
conditions. This formulation, based on a triple decomposition, paves
the way for understanding turbulence in such flows using only the mean
velocity profile.
Resolvent analysis based on the time-mean flow and dynamic mode
decomposition based on simulation data snapshots have both been used
to obtain a description of the reorganization of the flow structures
caused by the transpiration. We show that the pipe flows dynamics are
dominated by a critical-layer mechanism and the waviness induced in
the flow structures plays a role on
the streamwise momentum balance by generating additional terms. 
\end{abstract}

\section{Introduction}
\label{sec:intro}

The design of efficient flow control strategies for wall-bounded
turbulent flows aimed at reducing drag or energy expenditure remains
one of the most relevant and challenging goals in fluid mechanics
\citep{spalart2011drag}. For instance, approximately half the
propulsive power generated by airliner engines is employed to overcome
the frictional drag caused by turbulent boundary layers; however, the
scope of this challenge is much broader than aeronautical
engineering. Transport of fluids by pipeline, ubiquitous in industry,
nearly always occurs in a turbulent flow regime for which the energy
requirements can be significant. This is even more critical for long
distance pipelines designed to transport water, petroleum, natural gas
or minerals suspensions over thousands of kilometres. Another example
is the interstage pumping in processing plants, which can consume a
significant fraction of the energy used in the overall
process. Consequently, efficient flow control strategies for turbulent
boundary layers could achieve a dramatic impact on modern economies
\citep*{kim2011physics,mckeon2013experimental}. In the present work, we
focus on turbulent pipe flows.

\citet{kim2011physics} indicated that designing efficient flow control
strategies requires a deep understanding of the physical mechanisms
that act in wall-bounded turbulent flows, especially the
self-sustaining near-wall cycle \citep*{kim1987turbulence,
  jimenez1999autonomous}.  Its driving mechanism is often understood in terms of the
interaction of fluctuations with the mean shear and operator
non-normality \citep{Schmid:2007, Alamo.Jimenez:2006}.  As a
consequence, recent turbulence reduction control strategies are often
analysed \citep{Kim.Bewley:2007, kim2011physics} or designed
\citep{Sharma.Morrison.McKeon.ea:2011} with reduction of non-normality
in mind.  However, recent work has revealed the importance of the
critical layer amplification mechanism in asymptotically high Reynolds
number flow solutions \citep*{Blackburn.Hall.Sherwin:2013} and in
wall-bounded turbulence \citep*{McKeonSharma2010}.  In this context,
recent experimental findings in high Reynolds number wall-bounded turbulent flows
highlight the relevance of other coherent structures that scale with
outer variables, with streamwise length scales of several integral
lengths.

These flow structures, known as very large-scale motions (VLSM), were
reported by \citet{kim1999very}, \citet*{guala2006large} and
\citet{monty2007large}, who found that VLSM consist of long
meandering narrow streaks of high and low streamwise velocity that
contain an increasingly significant fraction of the turbulent kinetic
energy and shear stress production with increasing Reynolds
number. Thus the contribution of these flow structures to the overall
wall drag will be of the utmost importance at very high Reynolds
number. \citet{hutchins2007evidence} observed that these VLSM can
reach locations near the wall, thus flow control strategies applied to
the wall may have a strong influence on these motions. Therefore,
control of these VLSM structures may help achieve a drag increase or
reduction in high-Reynolds pipe flow. \citet{McKeonSharma2010} and
\citet{sharma2013coherent} showed that the sustenance of these
structures may in part be attributed to the critical layer
amplification mechanism. Consequently, it is important to understand
the influence of flow actuation on the critical layer.
VLSM become energetically non-negligible, in the sense of producing a
second peak in the streamwise turbulence intensity, at friction
Reynolds number $\Ret>10^4$ \citep*{smits2011high}.  Even though
computational experiments are almost unaffordable at these Reynolds
numbers, the behaviour of these structures can be observed in pipe
flow experiments at moderate bulk-flow Reynolds numbers
$\Rey=12\,500$, as shown by the proper orthogonal decomposition of PIV
data carried out by \citet{hellstrom2011visualizing} and
\citet{hellstrom2014energetic}.

Transpiration control, which is the application of suction and blowing
at the wall, can be used effectively to manipulate turbulent flow. One
of the first applications of transpiration can be found in the seminal
work by \citet*{choi1994active} in which they developed an active,
closed-loop flow control strategy known as opposition control,
consisting of a spatially distributed unsteady transpiration at a
channel wall. Their transpiration was a function of the wall-normal
velocity at a location close the wall and they were the first to
demonstrate that a significant drag reduction can be achieved by a
zero net mass flux transpiration. \citet{sumitani1995direct}
investigated the effect of (open-loop) uniform steady blowing and
suction in a channel flow. They applied blowing at one wall and
suction at the other, and concluded that injection of flow decreases
the friction coefficient and activates near-wall turbulence, hence
increasing the Reynolds stresses, and that suction has the opposite
effect. \citet{jimenez2001turbulent}, in their numerical investigation
of channel flows with porous surfaces, considered a flow control
strategy based on active porosity, subsequently converted to an
equivalent to static transpiration, and showed that the near-wall
cycle of vortices/streaks can be influenced by the effect of
transpiration. Furthermore, \citet{min2006sustained} showed that
sustained sub-laminar drag can be obtained in a turbulent channel flow
by applying a travelling sinusoidal (varicose) transpiration at
certain frequencies.

\cite{luchini2008acoustic} pointed out that steady streaming induced
by the transpiration plays a major role in the drag reduction
mechanism.
\cite{hoepffner2009pumping} performed numerical simulations of a
channel with travelling waves in the axial direction of wall
deformation and travelling waves of blowing and suction in the
streamwise direction. They discovered that the streaming induced by
wall deformation induces an increased flow rate while that produced by
travelling waves of blowing and suction generate a decrease in flow
rate (the streaming flow is however not the only contributor to
overall flow rate). \citet*{quadrio2007effect} performed a parametric
investigation of low-amplitude steady wall transpiration in turbulent
channel flows, finding both drag-increasing and drag-reducing
configurations. They observed that while the frictional drag was
dramatically increased at small wavenumbers, a reduction in drag was
possible above a threshold wavenumber, related to the length scales of
near-wall structures. The drag modifications were explained by two
physical mechanisms: interaction with turbulence, consisting of a
reduction in turbulence fluctuations by extracting turbulent fluid and
blowing laminar fluid, and generation of a steady streaming opposite
to the mean flow. \cite*{woodcock2012induced} carried out a
perturbation analysis of travelling wall transpiration in a
\twod\ channel, finding that the flux induced by the streaming opposes
the bulk flow. They conjectured that for three-dimensional flows and
beyond a certain transpiration amplitude, the transpiration effects
will only depend on the wavespeed.

In the present work we examine the effect of high- and low-amplitude
transpiration in turbulent pipe flow via direct numerical simulation
(DNS) at a moderate bulk flow Reynolds number $\Rey=10\,000$,
corresponding to friction Reynolds number $\Ret=314$. We focus on the
effect of steady wall-normal blowing and suction that varies
sinusoidally in the streamwise direction and with both high and low
transpiration amplitudes. The dataset consists of a wall transpiration
parameter sweep in order to assess the effect of the transpiration
parameters on the turbulence statistics and identify drag increasing
and reducing pipe configurations, as well as to permit quantitative
comparisons with previous trends observed in channel flows by
\citet{quadrio2007effect}.  Although the present values
of Reynolds number are not sufficient for a clear separation between
all the scales, we will draw special attention to the influence of the
flow control on VLSM-like structures. As shown by \citet{hellstrom2011visualizing}, motions corresponding to these large flow structures can be observed even at the considered bulk flow Reynolds number.

A resolvent analysis \citep{McKeonSharma2010} will be employed to
obtain the flow dynamics which are most amplified, with and without steady transpiration. This model-based framework consists of a gain analysis
of the \NavSto\ equations in the wavenumber/frequency domain, which
yields a linear relationship between the fluctuating velocity fields
excited by the non-linear terms sustaining the turbulence. This linear
operator depends on the mean profile, which is in turn sustained by
the Reynolds stresses generated by the fluctuations. This framework
has been already successfully employed in flow control by
\cite*{luhar2014opposition}, who modeled opposition control targeting
near-wall cycle structures. \citet{sharma2013coherent} employed the
same framework to recreate the behaviour of complex coherent
structures, VLSMs among them, from a low-dimensional subset of
resolvent modes. In the present context, the analysis permits the identification
of the flow structures that are amplified/damped by the effect of wall
transpiration and how their spatial functions are distorted by the
transpiration. We expand on this information in \S\ref{sec:resolanalysis} of the following.

The adoption of this critical layer framework in turn leads to an
analysis of the flow in the Fourier domain. Dynamic mode decomposition
(DMD) \citep{Schmid2010,RowleyEtAlJFM2009}, which works to pick out the dominant frequencies via snapshots
from the DNS data set, is its natural counterpart. In \S\ref{sec:6}, a DMD analysis on
the simulation data will also be carried out to identify the most energetic
flow structures at a given frequency and provide additional insight
into the flow dynamics.

\section{Direct numerical simulations}
\label{sec:DNS}

\begin{figure}
\centerline{\includegraphics*[width=0.8\linewidth]{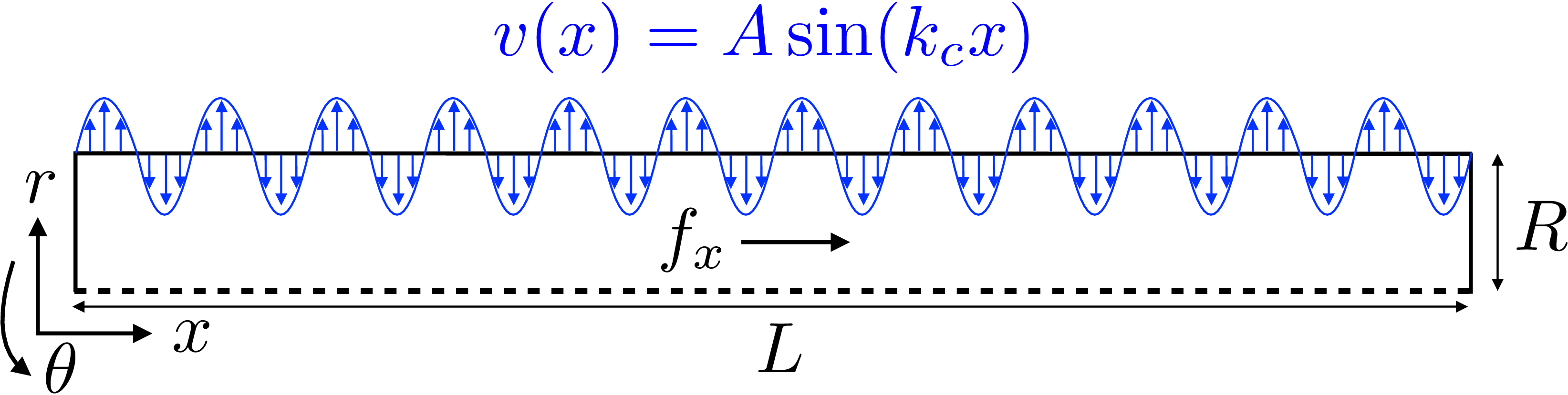}}
\caption{Physical domain and transpiration boundary condition}
\label{fig:geo}
\end{figure}

A spectral element--Fourier direct numerical simulation (DNS) solver
\citep{hugh2004} is employed to solve the incompressible
\NavSto\ equations in non-dimensional form
\begin{eqnarray}
\bm{\nabla \cdot \hat{u}} & = & 0
\\ \partial_t\bm{\hat{u}}+\bm{\hat{u}\cdot\nabla\hat{u}} & = &
-\bm{\nabla}p + \Rey^{-1}\nabla^2\bm{\hat{u}} + \bm{f}
\label{eqn:NSE}
\end{eqnarray} 
where $Re=U_b D/\nu$ is the Reynolds number based on the bulk mean velocity $U_b$, the pipe diameter $D$ and a constant kinematic viscosity $\nu$,
$\bm{\hat{u}}=(u,v,w)$ is the velocity vector expressed in cylindrical
coordinates $(x,r,\theta)$, $p$ is the modified or kinematic pressure,
 $\bm{f}=(f_x,0,0)$ is a forcing vector. A pipe with a periodic
domain of length $L=4\upi{R}$, where $R$ is the pipe's outer radius,
has been considered. No-slip boundary conditions for the streamwise and azimuthal velocity are applied at the pipe wall; transpiration in the wall-normal direction is applied to the flow by imposing the boundary condition,
\begin{equation}
v(x,R,\theta)=A\sin(k_c x) \, ,
\label{eq:bc}
\end{equation}
which represents steady sinusoidal wall-normal flow transpiration
along the streamwise direction with an amplitude $A$ and a streamwise
wavenumber $k_c$.  Additionally, $k_c$ must be an integer multiple of
the fundamental wavenumber in the axial direction $2\upi/L$ to enforce
a zero net mass flux over the pipe wall. A sketch of the
configuration is shown in figure~\ref{fig:geo}. The constant
streamwise body force per unit mass $f_x$ is added in (\ref{eqn:NSE})
to ensure that the velocity and pressure are streamwise periodic. 
(A simple physical equivalent is a statistically steady flow of liquid
driven by gravity in a vertical pipe which is open to the atmosphere
at each end.)
The body force $f_x$ is calculated on the basis of a
time-average force balance in the streamwise direction between the
body force exerted on the volume of fluid in the pipe and the traction
exerted by the wall shear stress, thus
\begin{equation}
\rho f_x L \upi R^2 = \tau_w 2 \upi R L \, ,
\label{eq:balance}
\end{equation}
with $\tau_w$ being the mean wall shear stress. Equivalently, it can
be shown that
\begin{equation}
\frac{f_x
  R}{2U_b^2}=\left(\frac{u_\tau}{U_b}\right)^2
  =\left(\frac{\Ret}{\Rey}\right)^2,
\label{eq:balance2}
\end{equation}
where $u_\tau=(\tau_w/\rho)^{1/2}$ is the friction
velocity, and $\Ret=u_\tau R/\nu$ is defined as the friction Reynolds number.
The low-$\Rey$ Blasius correlation \citep{blasius1913law} for
turbulent flow in a smooth pipe
\begin{equation}
\Ret=99.436 \times 10^{-3}\Rey^{7/8} \, ,
\label{eq:blasius}
\end{equation}
is employed to estimate the body force $f_x$ from
(\ref{eq:balance2}). In the present work, $\Ret=314$ was set on the
basis that for the zero-transpiration case, $\Rey=10\,000$;
consequently while $\Ret$ and $f_x$ are constants for the remainder of
this examination, $\Rey$ takes on different values for different
transpiration parameters. This is a direct indication of the drag reducing/increasing effect resulting from transpiration.  In order to alleviate this difference in $\Rey$, we will employ an alternative outer scaling with the Reynolds number based on the smooth pipe bulk mean velocity $U_b^s$,
\begin{equation}
\Rey_s = \frac{ U^s_b D}{\nu} 
\end{equation}
independently of whether the transpiration is applied or not. The fact that $\Rey_s$ is constant will be exploited later in comparing controlled and uncontrolled pipe flows via the streamwise momentum equation.

The spatial discretization employs a \twod\ spectral element mesh in a
meridional cross-section and Fourier expansion in the azimuthal
direction, thus the flow solution is written as
\begin{equation}
    \bm{\hat{u}}(x,r,\theta,t)= \sum_{\pm n} \bm{\hat{u}}_n
    (x,r,t)\ce^{\ci n\theta} \, .
\label{eq:FourierDecomp}
\end{equation}
Note that the boundary conditions in (\ref{eq:bc}) preserve
homogeneity in the azimuthal direction, so this Fourier decomposition
still holds for a pipe with wall-normal transpiration. The time is
advanced employing a second-order velocity-correction method developed
by \cite*{karniadakis1991high}. The numerical method, including
details of its spectral convergence in cylindrical coordinates, is
fully described in \citet{hugh2004}.  The solver has been previously
employed for DNS of turbulent pipe flow by \citet{chin2010influence},
\citet{saha2011influence}, \citet{saha2014scaling},
\citet{saha2015comparison}, \cite{SKOB-JFM-2015}, and validated
against the $\Ret=314$ smooth-wall experimental data of
\citet{den1997reynolds} in \citet{boc07}.
We use a mesh similar to that employed for the straight-pipe case of
\citet{SKOB-JFM-2015}, also at $\Ret=314$. The grid consists of 240
elements in the meridional semi-plane with a 11th-order nodal shape
functions and 320 Fourier planes around the azimuthal direction,
corresponding to a total of approximately $1.1\times10^7$
computational nodes. For transpiration cases in which the flow rate is
significantly increased, a finer mesh consisting of $1.6\times10^7$
degrees of freedom has been additionally employed. Simulations are
restarted from a snapshot of the uncontrolled pipe flow, transient
effects are discarded by inspecting the temporal evolution of the
energy of the azimuthal Fourier modes derived from
(\ref{eq:FourierDecomp}) and then statistics are collected until
convergence. Typically, 50--100 wash-out times ($L/U_b$) equivalently
to approximately 5000--10\,000 viscous time units are required for
convergence of the statistics.

In what follows, we will use either the \NavSto\ equations  (\ref{eqn:NSE}) non-dimensionalized with the smooth smooth pipe bulk velocity $U^s_b$, i.e with Reynolds number $Re_s$ independently of the transpiration, or non-dimensionalized with wall
scaling. This viscous scaling is denoted with a $+$ superscript.

\section{Flow control results}
\label{sec:3} 

\begin{figure}
  \centerline{\includegraphics*[width=1\linewidth]{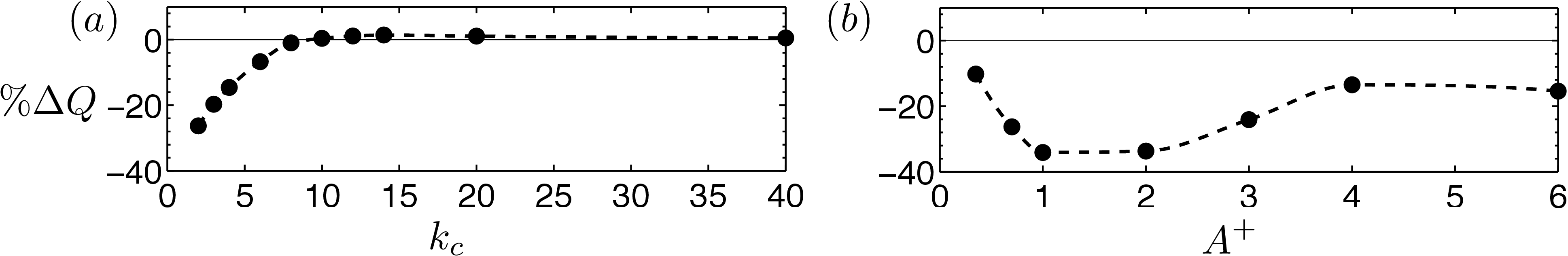}}\caption{Transpiration effect on the flow rate by changing (\textit{a})
  wavenumber $k_c$ at constant amplitude $A^+=0.7$ (\textit{b}) amplitude $A^+$
  at constant wavenumber $k_c=2$.}
\label{fig:Q}
\end{figure}

The independent effects of the two transpiration parameters, $A$ and
$k_c$, are investigated first. The range of parameters are similar to
the one employed by \citet{quadrio2007effect} in a
channel. Simulations consisting of a parameter sweep while maintaining
the other fixed have been carried out at $\Ret=314$. Following the
classical Reynolds decomposition, the total velocity has been
decomposed as the sum of the mean flow $\bm{u}_0$ and a fluctuating
velocity $\bm{u}$, which reads
\begin{equation}
{\bm {\hat{u}}}(x,r,\theta,t)= {\bm u}_0(x,r) + {\bm u}(x,r,\theta,t) \, ,
\label{def1}
\end{equation}
with the mean flow obtained by averaging the total flow in time and
the azimuthal direction as
\begin{equation}
\bm{u}_0(x,r) = \lim_{T \to \infty} \frac{1}{T} \int_0^T
\frac{1}{2\upi} \int_0^{2\upi} {\bm{\hat{u}}}(x,r,\theta,t)
\cd t \cd \theta \, .
\label{def2}
\end{equation}
For simplicity in the notation in what follows, we will use
$\langle~\rangle$ to denote an average in time and azimuthal
direction. Hence, $\bm{u}_0(x,r) = \langle{\bm{\hat{u}}} \rangle $ .

Note that the streamwise spatial dependence of the mean flow permits a
non-zero mean in the wall normal direction, hence
$\bm{u}_0(x,r)=(u_0,v_0,0)$. Turbulence statistics additionally
averaged in the streamwise direction are denoted with a bar
\begin{equation}
{{\bar{\bm u}_0}}(r) = \frac{1}{L}\int_0^L {\bm u}_0(x,r) \cd x \, .
\label{def3}
\end{equation}
In terms of flow control effectiveness, here we define drag-reducing or
drag-increasing configurations as those that reduce or increase the
streamwise flow rate with respect to the smooth pipe. Mathematically,
\begin{equation}
 \Delta Q=\frac{\int_0^R {\Delta{\bar{u}_0}} r \cd r}{\int_0^R {\bar{u}^s_0} r \cd r} \left\{
    \begin{array}{ll}
      < 0 & \mbox{drag-increasing}\, ,\\[2pt]
       >0 & \mbox{drag-reducing}\, ,
    \end{array} \right.
\label{eq:Q}
\end{equation}
where $\Delta{u_0}= (\bar{u}_0^c - \bar{u}_0^s)$, being $c$ and $s$
superscripts to denote controlled and smooth pipe respectively.
Figure \ref{fig:Q}(\textit{a}) shows the percentage variation of the flow rate
induced by transpiration with different wavenumber $k_c$ at constant
amplitude $A^+=0.7$, equivalent to $0.22\%$ of the bulk velocity.  A
large drag increase is observed at small wavenumbers and it
asymptotically diminishes until a small increase in flow rate is
achieved for $k_c \simeq 9$. This small drag reduction slowly lessens
and it is observed up to $k_c=40$.

Figure \ref{fig:Q}(\textit{b}) shows the transpiration influence on the flow
rate by increasing the amplitude $A^+$ at a constant wavenumber $k_c=2$.
A drag increase from zero transpiration up to
$A^+=4$ is observed. The maximum increase is achieved between
$A^+=1$ and $A^+=2$. In agreement with the conjecture posed by
\cite{woodcock2012induced}, increasing the amplitude beyond $A^+=4$
does not significantly change the value of the drag increase.

\subsection{Comparison with channel flow}

\begin{figure}
  \centerline{\includegraphics*[width=0.78\linewidth]{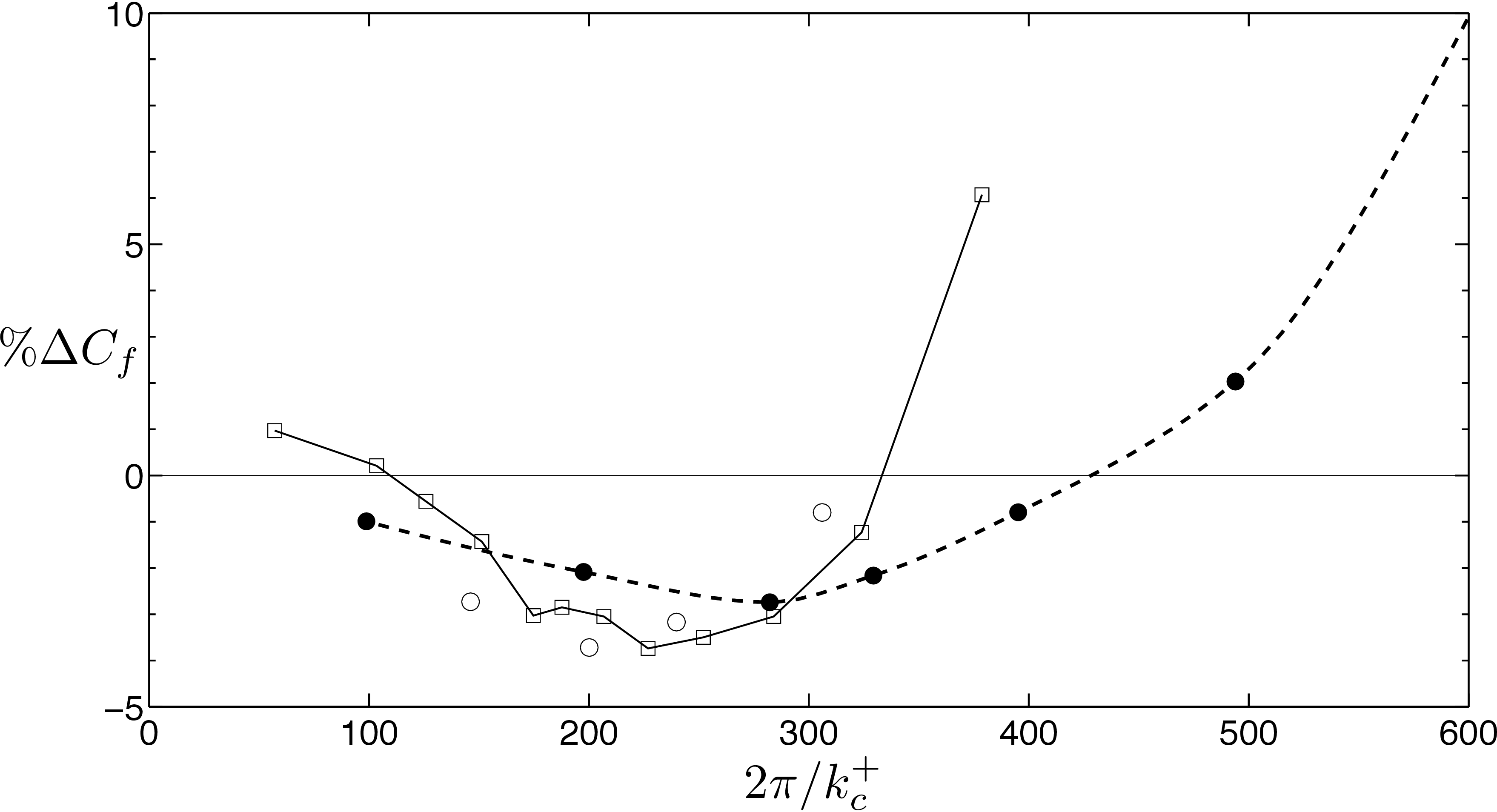}}
\caption{Comparison of the variation of the friction coefficient
  versus transpiration wavelength at fixed amplitude $A^+=0.7$. Solid
  circles, pipe flow at $Re_\tau=314$ (present results); open circles,
  channel flow at $Re_\tau=400$ \citep{quadrio2007effect}; squares,
  channel flow at $Re_\tau=180$ \citep{quadrio2007effect}.}
\label{fig:Quadrio}
\end{figure}

We compare our flow control results with the steady streamwise
transpiration study of \citet{quadrio2007effect} in a channel. The
variation of the mean friction coefficient $\Delta C_f$
with respect to the smooth pipe is
employed to assess the drag reduction/increase. The mean friction coefficient is defined as
\begin{equation}
C_f = \frac{2 \tau_w}{\rho U_b^2} \, .
\end{equation}
 Given that the mean wall shear stress is fixed, the mean friction coefficient
variation can be expressed solely as function of the ratio between the smooth and
controlled pipe bulk velocities, hence
\begin{equation}
\Delta C_f= \frac{1}{(1+\Delta Q)^2} -1  \, .
\end{equation}
in which the definition of $\Delta Q$ in (\ref{eq:Q}) is employed.  Note that negative values of $\Delta C_f$ correspond to drag reduction. Figure
\ref{fig:Quadrio} shows the comparison of the variation of the
friction coefficient versus transpiration wavelength at fixed
amplitude $A^+=0.7$ in pipe and channel flow. Low and high-$Re$
results are presented for the channel case. Similar behaviour is
observed, in which a drag reduction can be achieved in both channel and
pipe flows. The maximum friction variation is smaller for the pipe
than for the channel. In addition, the rate at which the friction
coefficient increases with wavelength is higher for the channel
than for the pipe. We note that the manner in which parameters are
varied differs between the two studies: while
\citet{quadrio2007effect} fixed the bulk velocity to arrange a
constant bulk Reynolds number during the parameter sweep, here we maintained
a constant friction Reynolds number $Re_\tau$. It is also worth noting
that because of the obvious difference in geometry, identical results
are not expected.

\section{Turbulence statistics \label{sec:4}}
\label{sec:turb}

\subsection{Constant transpiration amplitude}

\begin{figure}
  \centerline{\includegraphics*[width=1\linewidth]{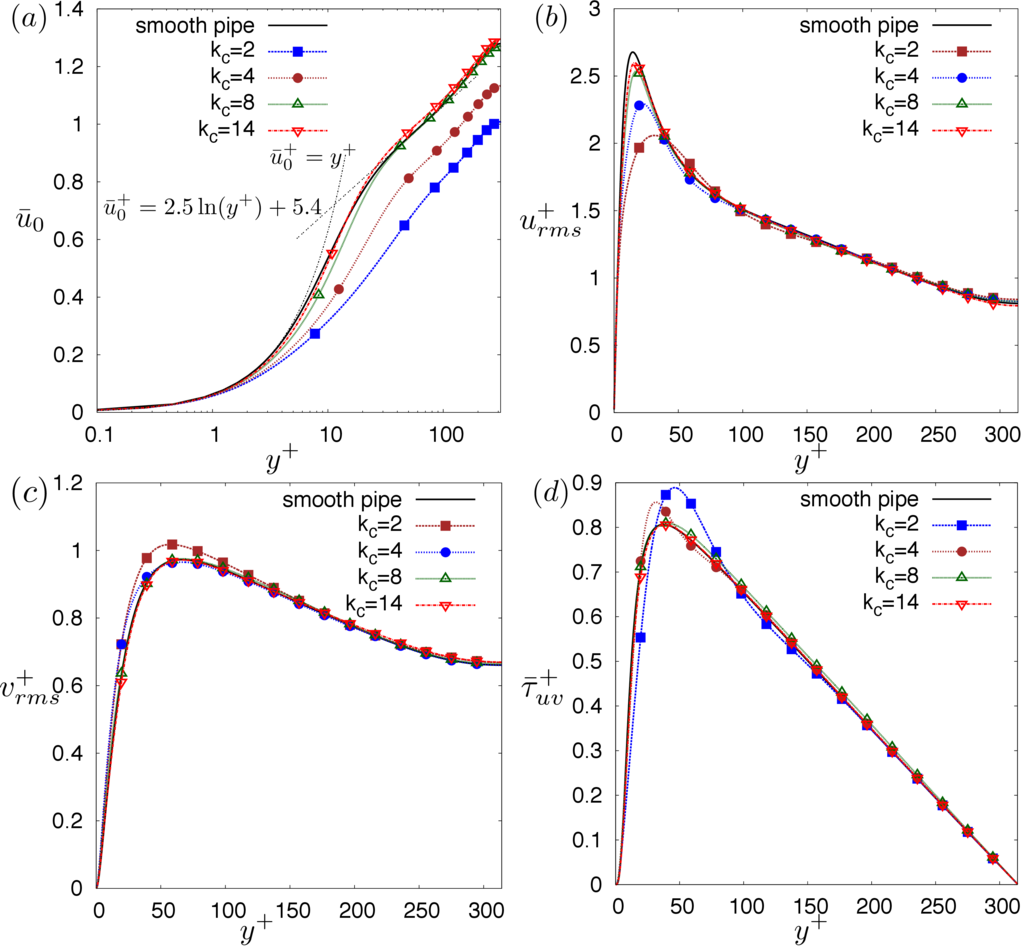}} \caption{Comparison of profile data for the smooth-wall pipe (solid
  line) and pipe with transpiration at constant amplitude $A^+=0.7$
  and different wavenumbers $k_c$. (\textit{a}), mean flow normalized
  by $U_b$, with dashed lines for linear sublayer and fitted log law;
  (\textit{b}), axial turbulent intensity; (\textit{c}), radial
  turbulent intensity; (\textit{d}), Reynolds shear stress.  }
\label{fig:UB_k}
\end{figure}

The effect of transpiration is evident in the behaviour of the mean
velocity characteristics. Figure \ref{fig:UB_k}(\textit{a}) shows the
effect of changing the transpiration wavenumber $k_c$ at a small
constant amplitude $A^+=0.7$ on the mean streamwise velocity
$\bar{u}_0(r)$ compared to the smooth pipe. Additionally, dashed lines
for the linear sublayer and fitted log law \citep{den1997reynolds} are
shown as reference. In agreement
with figure \ref{fig:Q}(\textit{a}), we observe reduced flow rates in
figure~\ref{fig:UB_k}(\textit{a}) for $k_c < 8$ with a small increase
occurring around $k_c=10$. Two interesting features
are noticed. First, the viscous sublayer is no longer linear. Since
the mean wall-shear $\tau_w$ is constant, all the profiles must
collapse as $y^+=(R^+-r^+)$ approaches zero. In other words, the value
of ${\partial u_0}/{\partial y}$ at the wall is the same for all cases
considered.

\begin{figure}
  \centerline{\includegraphics*[width=0.7\linewidth]{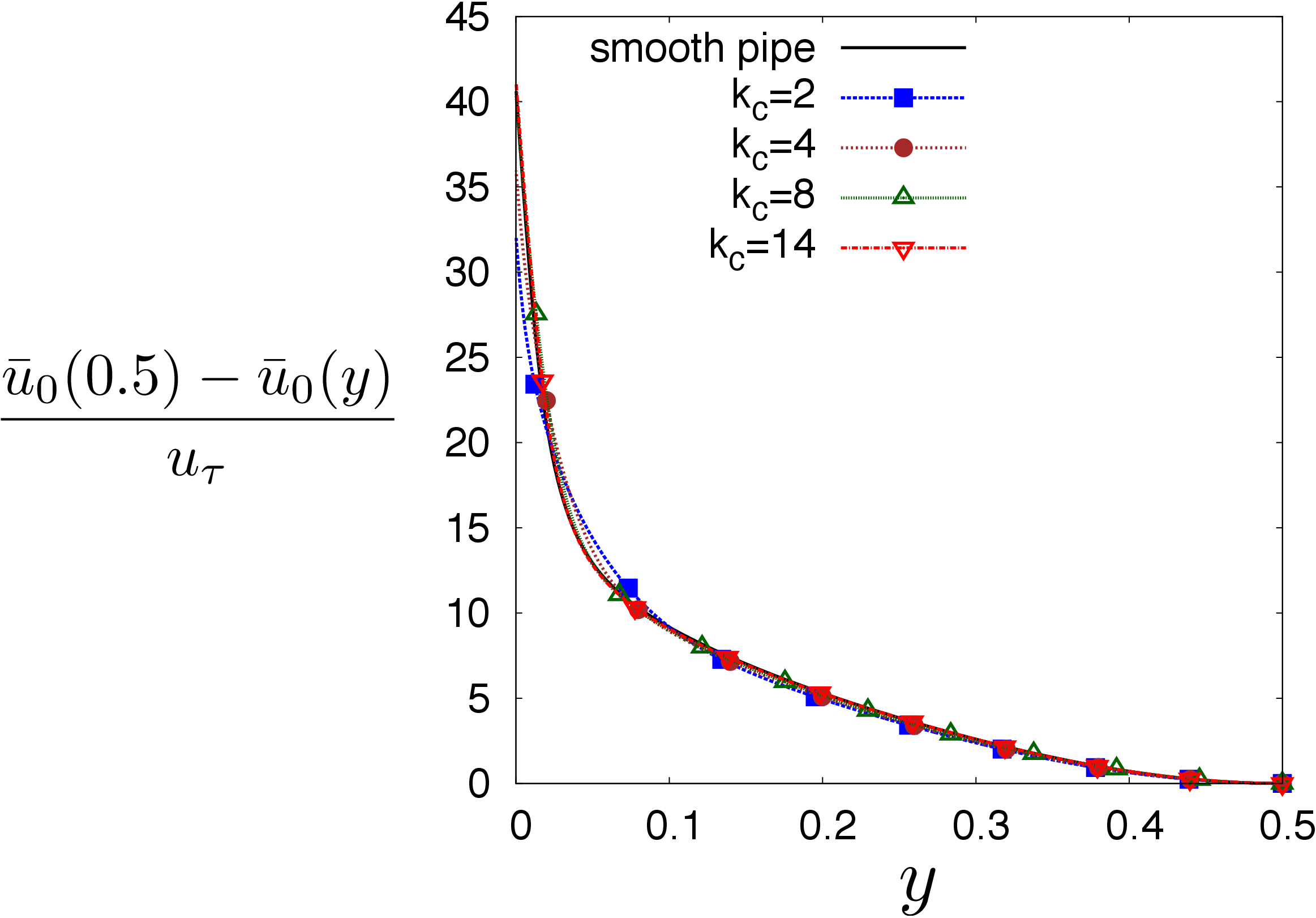}}
\caption{Defect velocity law for profile data of the smooth-wall pipe
  (solid line) and pipe with transpiration at constant amplitude
  $A^+=0.7$ and different wavenumbers $k_c$. A collapse of all the mean
velocity profiles is observed in the outer layer hence small amplitude effects do not alter the turbulent activity in the outer layer.} 
\label{fig:defect}
\end{figure}

Second, the outer-layer of all the profiles are
parallel, suggesting that the overlap region can be expressed as
\begin{equation}
\bar{u}_0^+=2.5\ln(y^+) + 5.4 + \Delta T^+ \, ,
\end{equation}
with the transpiration factor,  $\Delta T$ arising in a similar way as the
roughness factor of \citet{hama1954boundary}, or the corrugation
factor of \citet{SKOB-JFM-2015}.  However, we note that the present Reynolds number may not be sufficiently high for the emergence of a self-similar log law in the overlap region. Although there is not a well-defined log layer at such low $\Rey$, we could refer to it as log layer for convenience. The present results suggest what will happen at a greater $\Rey$. Figure \ref{fig:defect} shows a
defect velocity scaling of the mean profiles. A collapse of the mean
velocity profiles is observed in the outer layer, hence Townsend's wall similarity hypothesis \citep{Townsend1976} applies in the present low amplitude transpiration cases.

Figure \ref{fig:UB_k}(\textit{b}) and (\textit{c}) show the
influence of the transpiration on the turbulent intensities, presented as root mean squares. At this
small amplitude, the transpiration mainly affects the location and
value of the maximum turbulent intensities and all profiles
collapse as $y^+$ approaches the centerline, indicating that small
amplitude transpiration does not alter the turbulent activity in the outer
layer. Note that the peak in axial turbulent intensity moves due to
the reduction in shear at the same wall-normal location in all the
cases, which is apparent in the man velocity profiles in Fig. \ref{fig:UB_k}(\textit{a}).

Figure \ref{fig:UB_k}(\textit{d}) shows that the influence of small
amplitude transpiration on the Reynolds shear stress $\tau_{uv} =
\langle u^c v^c \rangle $. We observed that the transpiration
influence on $\tau_{uv}$ is only significant at small wavenumbers and,
as with the turbulent intensities, small amplitude transpiration does
not change the outer layer of the profile. The small changes in the
shear stress are not enough to explain the significant variations in
the mean flow, hence it is inferred that additional effects related to
steady streaming and non-zero mean streamwise gradients play a major
role in the momentum balance.

\begin{figure}
  \centerline{\includegraphics*[width=1\linewidth]{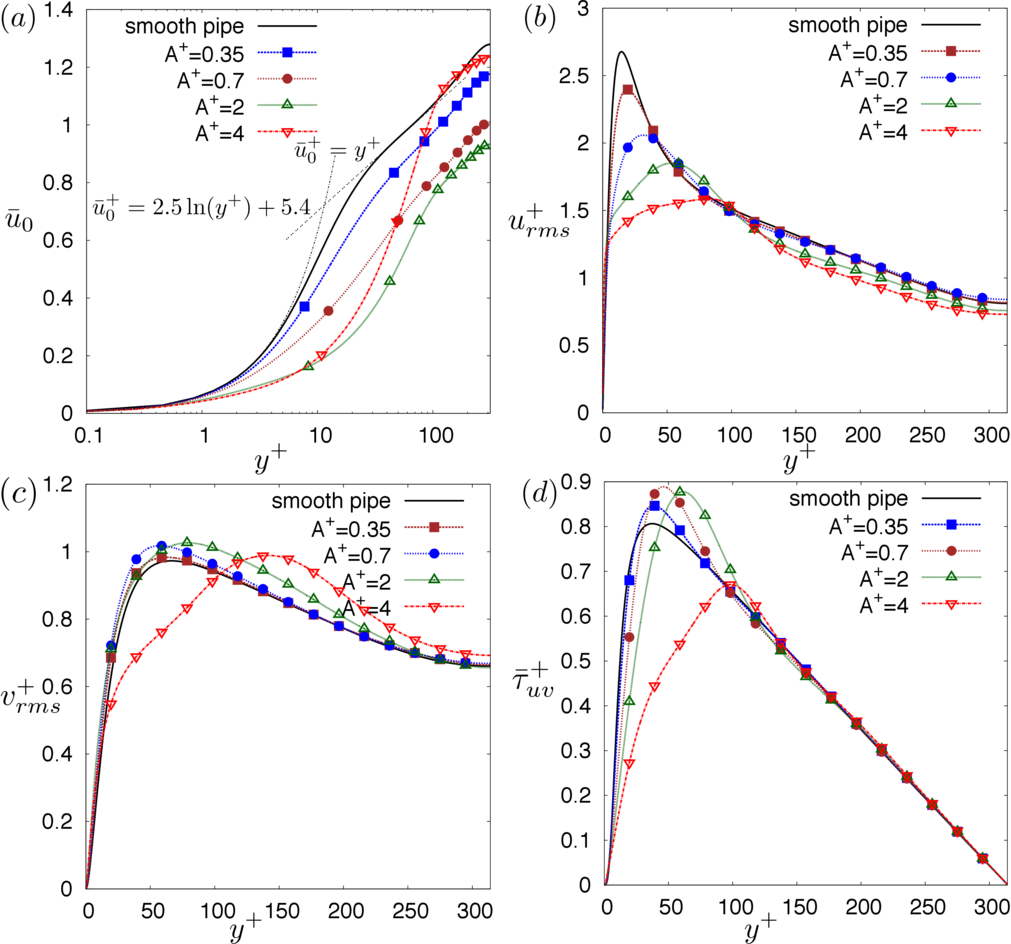}}
  \caption{Comparison of profile data for the smooth-wall pipe (solid
    line) and pipe with transpiration at constant wavenumber $k_c=2$
    and different amplitudes $A^+$. (\textit{a}), mean flow normalized
    by $U_b$, with dashed lines for linear sublayer and fitted log
    law; (\textit{b}), axial turbulent intensity; (\textit{c}), radial
    turbulent intensity; (\textit{d}), Reynolds shear stress.  }
\label{fig:UB_A}
\end{figure}

\subsection{Constant transpiration wavenumber}

The effect of changing the amplitude $A^+$ at a constant forcing
wavenumber $k_c=2$ in the mean streamwise velocity $u_0$ is shown in
figure~\ref{fig:UB_A}(\textit{a}). For $A^+<2$, results are similar to those observed for small amplitudes.  At $A^+=2$ the flow rate is dramatically decreased and the overlap region
is reduced compared to the small amplitude cases. At $A^+=4$ the
parallel overlap region is not existent. This large amplitude can
significantly increase the mean velocity in the outer
layer. Similarly, we observe that large amplitude transpiration shifts the location
of maximum turbulent intensities towards the centerline, as seen in
Figures \ref{fig:UB_A}(\textit{b}) and (\textit{c}). We can state that large amplitudes
$A^+ > 2$ can influence the turbulent activity in the outer
layer. Figure \ref{fig:UB_A}(\textit{d}) shows that high amplitudes have a similar effect on the shear stress and how a high amplitude $A^+= 4$ can
dramatically reduce the Reynolds shear stress and shift the location
of the maximum towards the centerline.

These tendencies in $A^+$ and $k_c$ suggest that very high values of
the transpiration amplitude combined with high wavenumber $k_c$ can be
explored in order to find significant drag-reducing
configuration. This has been confirmed through the investigation of a
static transpiration case with an amplitude $A^+=10$ and wavenumber
$k_c=10$, yielding an increase in flow rate of $\Delta Q=19.3\%$. This
case will be investigated in detail in the next section.

\section{Streamwise momentum balance}
\label{sec:5}

As previously mentioned, we have inferred from figures~\ref{fig:UB_k}
and~\ref{fig:UB_A} that the changes in Reynolds stress are not enough
to explain the changes in the mean velocity profile; additional
effects related to steady streaming and non-zero mean streamwise
gradients could play a major role in the momentum balance. Here we analyze
the streamwise momentum balance in order to identify these additional
effects. The axial momentum equation averaged in time and azimuthal
direction for a pipe flow controlled with static transpiration is
\begin{equation}
f_x + \frac{1}{r}\frac{\partial}{\partial{r}}(-r\tau_{uv}^c - ru^c_0v^c_0 + r\Rey_s^{-1}\frac{\partial{u^c_0}}{\partial{r}}) + \mathcal{N}_x = 0 \, .
\label{controlled}
\end{equation}
with $\mathcal{N}_x$ being the sum of terms with $x-$derivatives,
representing the non-homogeneity of the flow in the streamwise
direction \citep{fukagata2002contribution}. 

\begin{equation}
\mathcal{N}_x = \frac{\partial(u_0^c u_0^c)}{\partial x} + \frac{\partial \langle u^c u^c \rangle }{\partial x} - \Rey_s^{-1}\frac{\partial^2 u_0^c}{\partial x^2} \,.
\end{equation}
where $\langle \rangle$ denotes averaging in time and the azimuthal
direction. Equation ({\ref{controlled}}) for a smooth pipe yields
\begin{equation}
 f_x + \frac{1}{r}\frac{\partial}{\partial{r}}(-r\tau_{uv}^s + r\Rey_s^{-1}\frac{\partial{u^s_0}}{\partial{r}}) = 0 \, ,
\label{smooth}
\end{equation}
Since the body force $f_x$ and Reynolds number $\Rey_s^{-1}$ have the same value in
(\ref{controlled}) and (\ref{smooth}), these two equations can be
subtracted and integrated in the wall normal direction to give
\begin{equation}
\Rey_s^{-1}{\Delta{u_0}(x,r)}=\int_0^r \Delta{\tau} r^\prime \mathrm{d}r^\prime + \int_0^r u^c_0v^c_0 r^\prime \mathrm{d}r^\prime + \int_0^r\mathcal{N}^\prime_x r^\prime \mathrm{d}r^\prime \, ,
\end{equation}
in which $\mathcal{N}_x$ has been previously integrated with respect
to the wall normal direction to yield $\mathcal{N}^\prime_x$. This
equation can be additionally averaged in the axial direction to
identify three different terms playing a role in the modification of
the mean profile

\begin{equation}
\Delta{\bar{u}_0}(r)=RSS+ST+NH \, ,
\label{balance}
\end{equation}
where
\begin{eqnarray}
RSS(r) &=& \frac{-\Rey_s}{L}\int_0^L\int_0^r \Delta{\tau} r^\prime \mathrm{d}r^\prime \mathrm{d}x\, , \\ 
ST(r) &=& \frac{-\Rey_s}{L}\int_0^L\int_0^r u^c_0v^c_0 r^\prime \mathrm{d}r^\prime\mathrm{d}x\, , \label{ST} \\ 
NH(r) &=&  \frac{\Rey_s}{L}\int_0^L\int_0^r\mathcal{N}^\prime_x r^\prime\mathrm{d}r^\prime \,  \mathrm{d}x\, .
\label{NH}
\end{eqnarray}
The first term $RSS$ represents the interaction of the transpiration
with the Reynolds shear stress, and its behaviour can be inferred from
the turbulence statistics shown in Figures \ref{fig:UB_k}(\textit{d})
and \ref{fig:UB_A}(\textit{d}). The second term $ST$ is associated
with the steady streaming produced by the static transpiration. As
briefly mentioned in the introduction, this term defined in (\ref{ST}) consists of an additional flow rate due to the interaction of the wall-normal transpiration with the flow
convecting downstream \citep{luchini2008acoustic}.  This term can be
alternatively understood as a coherent Reynolds shear stress. The
velocity averaged in time and azimuthal direction defined in
(\ref{def2}) can be decomposed into a mean profile  ${{\bar{\bm u}_0}}(r)$ and a
steady deviation from that mean velocity profile $\bm{u}_0^\prime(x,r)
$
\begin{equation}
\bm{u}_0(x,r)  = {{\bar{\bm u}_0}}(r) +  {\bm u}_0^\prime(x,r)  \, .
\label{eq:triple}
\end{equation}
Taking into account this decomposition, the integrand of the steady
streaming term now reads
\begin{equation}
u^c_0 v^c_0  = {{\bar{u}^c_0}} {{\bar{v}^c_0}}  +  {{\bar{u}^c_0}} {v^c_0}^\prime \,  +  {{\bar{v}^c_0}} {u^c_0}^\prime \, +  {u^c_0}^\prime {v^c_0}^\prime \,.
\end{equation}
and its axial average is reduced to
\begin{equation}
 \frac{1}{ L} \int_0^L u^c_0 v^c_0 \mathrm{d}x =  \frac{1}{ L} \int_0^L  {u^c_0}^\prime {v^c_0}^\prime  \mathrm{d}x \,,
\end{equation}
since ${{\bar{v}^c_0}}$ must be zero because of the continuity
equation. Hence the steady streaming term $ST$ is generated by
coherent Reynolds shear stress induced by the deviation of the
velocity from the axial mean profile. We highlight that the
substitution of the decomposition (\ref{eq:triple}) in the Reynolds
decomposition in (\ref{def1}) leads to a triple decomposition, in
which the deviation from the mean velocity profile plays the role of
a coherent velocity fluctuation.

The third term $NH$ defined in (\ref{NH}) corresponds to the axial non-homogeneity in the
flow induced by the transpiration. It consists of non-zero mean
streamwise gradients generated by the transpiration. Because of the
low amplitudes considered in previous works \citep{quadrio2007effect},
this last term has not been isolated before and has been implicitly
absorbed in an interaction with turbulence term that accounts for both
$ST$ and $NH$ terms.

Finally, we note that the identity derived by
\citet{fukagata2002contribution} (FIK identity) could be alternatively
employed for our analysis. However, the difference in bulk flow
Reynolds numbers between uncontrolled and controlled cases favors the
present approach.

\subsection{Representative transpiration configurations}

The relative importance of the three terms acting in the transpiration
is investigated by inspecting three representative transpiration
configurations in terms of drag modification. Table \ref{tab:cases}
lists the contributions to the change in flow rate of these configurations: (I) the largest drag-reducing case found
($\Delta Q=19.3\%$) consisting of a large amplitude and transpiration wavenumber $(A^+,k_c)=(10,10)$, (II) a small
drag-reducing case ($\Delta Q=0.4\%$) induced by
a tiny amplitude at a large wavenumber $(A^+,k_c)=(0.7,10)$, and (III) a large
drag-increasing case ($\Delta Q=-36.1\%$) with small wavenumber and
amplitude $(A^+,k_c)=(2,2)$.

\begin{figure}
\centerline{\includegraphics*[width=1\linewidth]{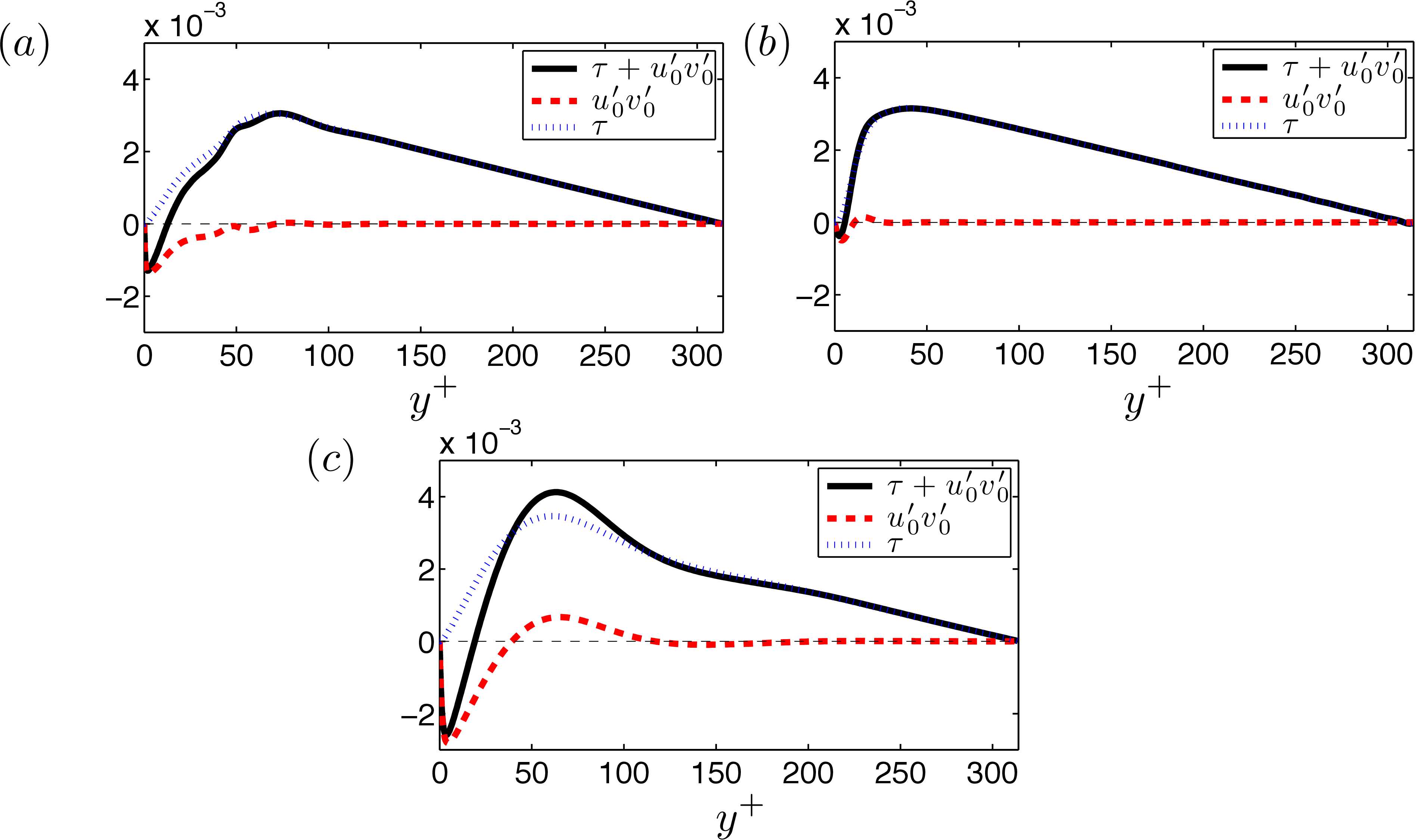}}
\caption{\label{fig:3RSS}Radial distribution of the Reynolds shear
  stress generated by the fluctuating velocity $\tau_{uv}$ and the
  deviation velocity ${u^c_0}^\prime {v^c_0}^\prime$ (\textit{a}) case I: large
  drag-decrease $(A^+,k_c)=(10,10)$, (\textit{b}) case II: small drag-decrease
  $(A^+,k_c)=(0.7,10)$, (\textit{c}) case III: drag-increase
  $(A^+,k_c)=(2,2)$.}
\end{figure}

We first compare the Reynolds shear stress generated by the
fluctuating velocity $\tau_{uv}$ and the shear stress arising from the deviation velocity ${u^c_0}^\prime {v^c_0}^\prime$. The radial distributions of
the three different cases are showed in figure~\ref{fig:3RSS}. We
observe that the streaming or coherent Reynolds shear stress is
dominant close to the wall and opposes the Reynolds stress generated
by the fluctuating velocity. Far from the wall, the Reynolds stress is
governed by the fluctuating velocity.

\begin{table}
  \begin{center}
\def~{\hphantom{0}}
  \begin{tabular}{c l c c r| r r r }
 &  & $A^+$ & $k_c$  & $ \Delta Q (\%)$ & $\int RSS (\%)$ & $\int ST(\%)$ & $\int NH(\%)$ \\ [3pt]
(I) & large drag-reduction & 10 & 10 & 19.3 &  28.4 &  $-20.4$  &  11.3   \\ 
(II)& small drag-reduction &  0.7 & 10 & 0.4&  1.5   &  $-1.7$  &  0.6  \\ 
(III)& drag-increase  & 2 & 2 & $-36.1$ &  11.2 &  $-29.6$ & $-17.7$  \\ 
  \end{tabular}
  \caption{Contributions to the change in flow rate of different transpiration configurations based on integration of Equation (\ref{balance})($\int$ denotes integration in wall normal direction).}
  \label{tab:cases}
  \end{center}
\end{table}

\begin{figure}
  \centerline{\includegraphics*[width=1\linewidth]{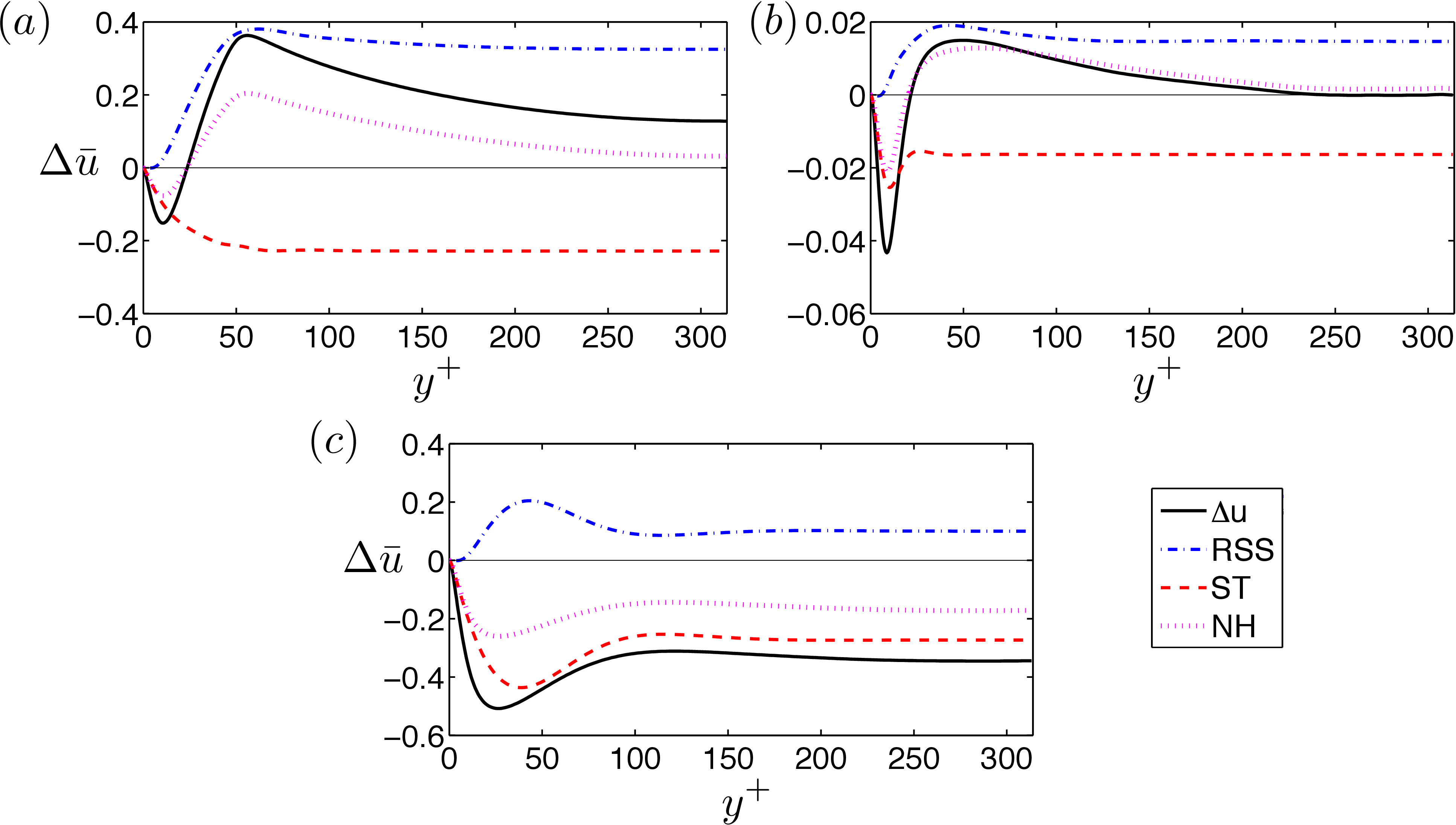}}
\caption{\label{fig:3cases}Radial distribution of the different terms
  in (\ref{balance}) to the streamwise momentum balance. Solid line
  $\Delta{\bar{u}_0}(r)$; \textcolor{blue}{-$\cdot$-} Reynolds shear
  stress term$RSS$; \textcolor{red}{$--$} steady streaming terms $ST$;
  \textcolor{magenta}{$\cdot$$\cdot$$\cdot$$\cdot$} non-homogeneous
  terms $NH$. (\textit{a}) case I: large drag-decrease $(A^+,k_c)=(10,10)$, (\textit{b})
  case II: small drag-decrease $(A^+,k_c)=(0.7,10)$, (\textit{c}) case III:
  drag-increase $(A^+,k_c)=(2,2)$.}
\end{figure}

\begin{figure}
  \centerline{\includegraphics*[width=1\linewidth]{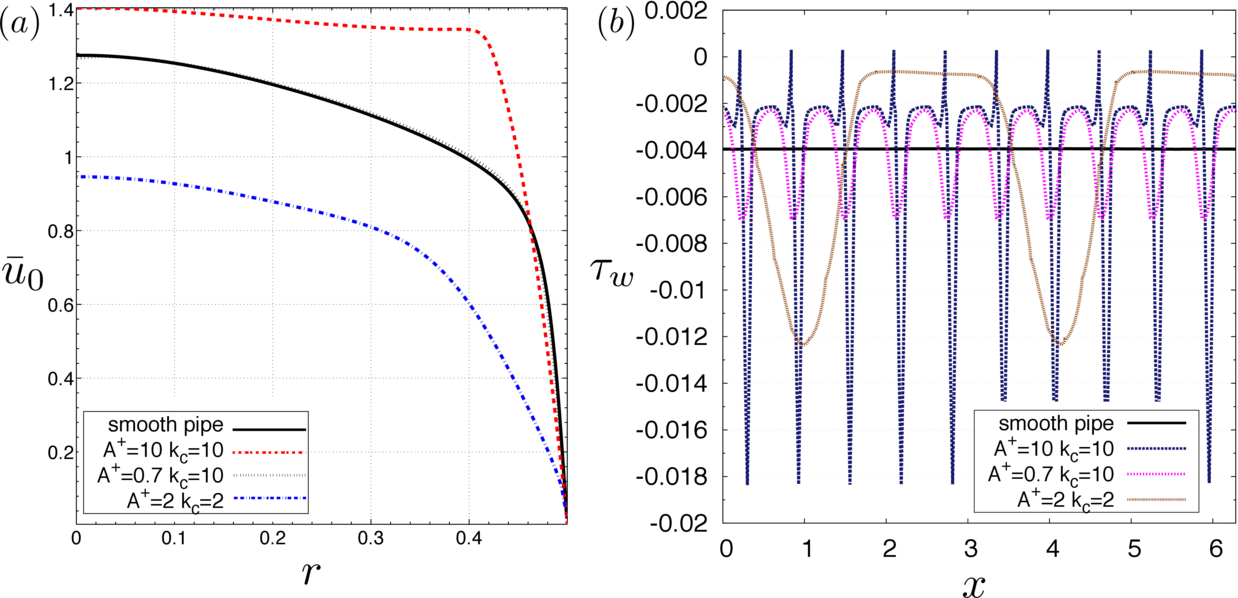}}
\caption{\label{fig:meanshear} (\textit{a}) Streamwise mean profile (\textit{b}) Mean wall shear along the axial direction}
\end{figure}

Figure~\ref{fig:3cases} shows the radial profile of the different
contributions to the mean profile and figure~\ref{fig:meanshear} the
corresponding mean profiles of the cases considered. Figure
\ref{fig:3cases}(\textit{a}) corresponds to the drag-decrease case
$(A^+,k_c)=(10,10)$. It can be observed that the contributions to the
increase in flow rate are caused by a large reduction of Reynolds
shear stress (28.4\%) and the streamwise gradients produced in the
flow at such large amplitude and wavenumber (11.3\%). However, these
contributions are mitigated by the opposite flow rate induced by the
steady streaming ($-20.4\%$). Figure~\ref{fig:3cases}(\textit{b}) corresponds to
a low amplitude static transpiration $A^+=0.7$ at the same axial
wavenumber as case (\textit{b}), yielding a very small reduction in flow
rate. Similarly, the effect on the Reynolds shear stress (1.5\%) and
the non-homogeneity effects (0.6\%) contribute to increase the flow
rate while the streaming produced is opposite to the mean flow, but of
a lesser magnitude ($-1.7\%$).  Figure \ref{fig:3cases}(\textit{c}) represents
the drag-increase case $(A^+,k_c)=(2,2)$. As opposed to the two
previous cases, the main contribution to the drag is caused by a large
steady streaming in conjunction with a large non-zero streamwise
gradients opposite to the mean flow. Although the decrease in Reynolds
stress is significant and favorable towards a flow rate increase, the
sum of the other two terms is much larger. In terms of the variation
of mean profile, we observe a similar behaviour in all cases: there is
a decrease in the mean streamwise velocity profile close to the wall induced by the
streaming term or coherent Reynolds shear stress. Within the buffer
layer, the difference in mean velocity decreases until a minimum is
achieved and then it increases to a maximum in the overlapping region. A smooth
decrease towards the centerline is then observed. 

\begin{figure}
\begin{center}
\includegraphics*[width=1\linewidth]{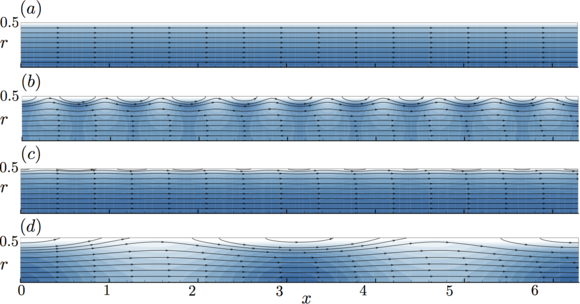}
\caption{Two-dimensional mean flow: streamlines and normalized kinetic
  energy for each different case (white to blue) ($a$) reference case: smooth
  pipe ($b$) drag-reducing case $(A^+,k_c)=(10,10)$ ($c$) neutral case
  $(A^+,k_c)=(0.7,10)$ ($d$) drag-increasing case
  $(A^+,k_c)=(2,2)$ \label{fig:means} }
\end{center}
\end{figure}

Streamlines and kinetic energy of the \twod\ mean velocities
$\bm{u}_0(x,r)$ are shown in figure~\ref{fig:means}. Preceding the
analysis of the flow dynamics, we observe that blowing is associated
with areas of high kinetic energy while suction areas slow down the
flow.  Despite the large amplitudes employed, we observe an apparent
absence of recirculation areas in these mean flows. Figure
\ref{fig:meanshear}(\textit{b}) shows the mean wall shear along the axial
direction. Only a very small flow separation occurs in the large
drag-decrease $(A^+,k_c)=(10,10)$. Hence flow separation effects
are not associated with the observed drag modifications.

\subsection{Flow control energetic performance}
\label{sec:energy}

As stated by \citet{quadrio2011drag}, a flow control system study must
always be accompanied with its respective energetic perfomance
analysis in order to determine its validity in real applications. The
control performance indices in the sense of \citet{kasagi2009toward}
are employed here to assess the energy effectiveness of the
transpiration. We define a drag reduction rate as
\begin{equation}
W = (P_c - P_s)/{P_s} \, ,
\label{eq:P}
\end{equation}
where $P$ is the power required to drive the pipe flow and the
subscripts $c$ and $s$ refer again to controlled and smooth pipe. The
index $W$ can also be interpreted as the proportional change in the power developed
by the constant body force $f_x$ as a consequence of the
transpiration. Note from (\ref{eq:P}) and (\ref{eq:pump}) that the drag reduction
rate $W$ is equivalent to the flow rate variation $\Delta Q$. The power required to drive the flow is given by the
product of the body force times the bulk velocity
\begin{equation}
P=f_x \upi R^2 L U_b \, .
\label{eq:pump}
\end{equation}  A net energy saving rate $S$ is defined to take into account the power
required to operate the flow control
\begin{equation}
S=\left(P_s - (P_c + P_{in})\right)/{P_s} \, .
\end{equation}
The mean flow momentum and energy equation are employed to obtain the power required to apply the transpiration control, as also done by \citet{marusic2007laminar} and \citet{Mao2015cylinder}. The power employed to apply the transpiration control reads
\begin{equation}
P_{in} =  \int_0^{2\upi} \int_0^{L} \left( \frac{1}{2}v_0(x,R)^3 + p(x,R)v_0(x,R) \right) \, \cd x R \cd \theta .
\end{equation}
with $v_0(x,R)=A\sin(k_c x)$. The first term in the integral represents the rate at which energy is introduced or removed as kinetic energy in the flow through the pipe wall. This term is exactly zero because the change of net mass flux is zero and the velocity is sinusoidal, imposed by the
transpiration boundary condition in (\ref{eq:bc}). The second term
represents the rate of energy expenditure by pumping flow against the
local pressure at the wall. Finally, the effectiveness of the
transpiration control is defined as the ratio between the change in
pumping power and power required to apply the transpiration control,
which reads
\begin{equation}
G=(P_s-P_c)/P_{in} \, .
\end{equation}
Note that the net power saving can be alternatively written as
\begin{equation}
S=W(1-G^{-1}) \, ,
\end{equation}
which indicates that an effectiveness $G$ higher than one is required
for a positive power gain.  As reported by
\citet{kasagi2009toward}, typical maxima for active feedback control
systems are in the range of $G\sim100$ and $S\sim0.15$ and for
predetermined control strategies, such as spanwise wall-oscillation
control in channel \citep{quadrio2004critical} or streamwise
travelling transpiration in channel \citep{min2006sustained}, a range
of $2\lesssim G\lesssim6$ and $0.05\lesssim S\lesssim0.25$ was found.

\begin{table}
  \begin{center}
\def~{\hphantom{0}}
  \begin{tabular}{l c c | r r r}
 & $A^+$ & $k_c$  & $ W$ & $G $ & $S $ \\ [3pt]
(I) large drag-reduction & 10 & 10 & 0.19&  1.02 & 0.004  \\ 
(II) small drag-reduction &  0.7 & 10 & 0.04  & 0.21 & -0.014 \\ 
(III) drag-increase  & 2 & 2 & -0.36  & -3.73  & -0.46 \\ 
  \end{tabular}
  \caption{Flow control energetic performance indices. $W$ represents the drag reduction rate, $G$ the effectiveness of the
transpiration control and $S$ the net energy saving rate.}
  \label{tab:FCPI}
  \end{center}
\end{table}

The resulting flow control energetic performance indices are listed in
table~\ref{tab:FCPI}. Note that the change in pumping power rate $W$
is equivalent to the flow rate variation $\Delta Q$. Despite the large
drag reduction produced by the configuration $(A^+,k_c)=(10,10)$, the
net energy saving rate is marginal and the small drag reduction caused
at $(A^+,k_c)=(0.7,10)$ has an effectiveness less than one, with a net
energy expenditure. The case $(A^+,k_c)=(2,2)$ is interesting, as it
shows a high effectiveness in decreasing the flow rate and has the
potential to relaminarize the flow at lower $Re_\tau$. For instance, a
reduction in flow rate $\Delta Q=-36\%$ at $Re_\tau=115$ could reduce
the critical bulk flow Reynolds below $Re<2040$ and thus achieve a
relaminarization of the flow. Finally, we remark that the particular
case $(A^+,k_c)=(10,10)$ is probably not the globally optimal
configuration for steady transpiration and net energy saving rates and
effectiveness in the range of typical values for the flow control
strategies reported by \citet{kasagi2009toward} might be possible, if the full parameter space was considered.

\section{Flow dynamics}
 \label{sec:6}

As a first step in establishing a relationship between changes in flow structures
and drag reduction or increase mechanisms, this section describes how the most
amplified and energetic flow structures are affected by the transpiration.

\subsection{Resolvent analysis}
\label{sec:resolanalysis}

The loss of spatial homogeneity in the axial direction is the main challenge
when incorporating transpiration effects into the resolvent model. One solution
is to employ the \twod\ resolvent framework of \citet{gomez_pof_2014}. This
method is able to deal with flows in which the mean is spatially
non-homogeneous.  The dependence on the axial coordinate $x$ is retained in the
formulation, in contrast with the classical \oned\ resolvent formulation of
\citet{McKeonSharma2010}.  However, the method leads to a singular value
decomposition problem with storage requirements of order
$\mathcal{O}(N_r^2N_x^2)$, with $N_r$ and $N_x$ being the resolution in the
radial and axial directions respectively, as opposed to the original formulation
which was of order $N_r^2$. Consequently, the \twod\ method is not
practical for a parameter sweep owing to the large computational effort
required.  In the following we present a computationally cheaper and simple
alternative based on a triple decomposition of the total velocity.

Based on the decompositions in (\ref{def1}) and (\ref{eq:triple}), the
total velocity is decomposed as a sum of the axial mean profile, a
steady but spatially varying deviation from the axial profile and a
fluctuating velocity
\begin{equation}
\label{eq:triple}
\bm{\hat{u}}(x,r,\theta,t)=\bar{\bm{u}}(r) +
  \bm{u}^\prime(x,r) + \bm{u}(x,r,\theta,t).
\end{equation}
The fluctuating velocity is expressed as a sum of Fourier modes. 
These are discrete since the domain has a fixed periodic length and is
periodic in the azimuth:
\begin{equation}
\bm{\hat{u}}(x,r,\theta,t)=\bar{\bm{u}}(r) + \bm{u}_l^\prime(r)
\ce^{\ci lx} + \sum_{(k,n,\omega) \neq (l,0,0) } \bm{u}_{k,n,\omega}(r)
\ce^{\ci(kx+n\theta-\omega t)} + \CC,
\end{equation}
where $k$, $n$ and $\omega$ are the axial, azimuthal
wavenumber and the temporal frequency, respectively.  Note than a
\cc\ must be added because $\bm{\hat{u}}$ is real.  

Without loss of generality,
we have assumed that the deviation velocity can be expressed as a
single Fourier mode with axial wavenumber $l$.  As a consequence of
the triadic interaction between the spatial Fourier mode of the
deviation velocity and the fluctuating velocity, there is a coupling
of the fluctuating velocity at $(k,n,\omega)$ with that at $(k\pm
l,n,\omega)$ (see Appendix A).  Similarly, the non-linear forcing
terms generated by the fluctuating velocity are written as
$\bm{f}_{k,n,\omega}=(\bm{u\cdot\nabla u})_{k,n,\omega}$.

Taking (\ref{eq:triple}) into account, it follows that the Fourier-transformed
\NavSto\ equation (\ref{eqn:NSE}) yields the linear relation
\begin{equation}
\bm{u}_{k,n,\omega}=\mathcal{H}_{k,n,\omega}\left(\bm{f}_{k,n,\omega}
+ \mathcal{C}_{k,n,\omega} \bm{u}_{k \pm l,n,\omega} \right) \,
\end{equation}
for $(k,n,\omega) \neq (0,0,0)$ and $(k,n,\omega) \neq (\pm l,0,0)$,
where $\mathcal{C}_{k,n,\omega}$ is a coupling operator representing
the triadic interaction between deviation and fluctuating
velocity. The triadic interaction can be considered as another unknown
forcing and it permits lumping all forcing terms as
\begin{equation}
\bm{g}_{k,n,\omega} = 
\bm{f}_{k,n,\omega} + \mathcal{C}_{k,n,\omega} \bm{u}_{k \pm l,n,\omega} \, ,
\label{coupling}
\end{equation}
hence the following linear velocity--forcing relation is obtained
\begin{equation}
\label{eq.exres}
\bm{u}_{k,n,\omega}=\mathcal{H}_{k,n,\omega}\bm{g}_{k,n,\omega} \, .
\end{equation}
The resolvent operator $\mathcal{H}_{k,n,\omega}$ acts as a transfer
function between the fluctuating velocity and the forcing of the
non-linear terms, thus it provides information on which combination of
frequencies and wavenumber are damped/excited by wall transpiration
effects. Equation (\ref{discrtH}) shows the resolvent
$\mathcal{{H}}_{k,n,\omega}$. The first, second and third rows
corresponds to the streamwise, wall-normal and azimuthal momentum
equation, respectively. 

\begin{equation}
\setlength{\arraycolsep}{0pt}
\renewcommand{\arraystretch}{1.3}
 \mathcal{{H}}_{k,n,\omega}(r) =\left[
\begin{array}{ccc}
\ci (k u_0 - \omega) - \Rey^{-1}(D + r^{-2})   &  \partial_r{u_0} & 0 \\
  0 & \ci (k u_0 - \omega) - Re^{-1}D     & -2\ci nr^{-2}\Rey^{-1} \\
  0 & -2\ci nr^{-2}\Rey^{-1}   & \ci (k u_0 - \omega) - \Rey^{-1}D  \\
\end{array}  \right]^{-1} ,
\label{discrtH}
\end{equation}
with 
\begin{equation}
D=-k^2 - (n^2 +1)r^{-2} + \partial^2_r + r^{-1}\partial_r \, .
\end{equation}

\begin{figure}
\begin{center}
\includegraphics*[width=0.6\linewidth]{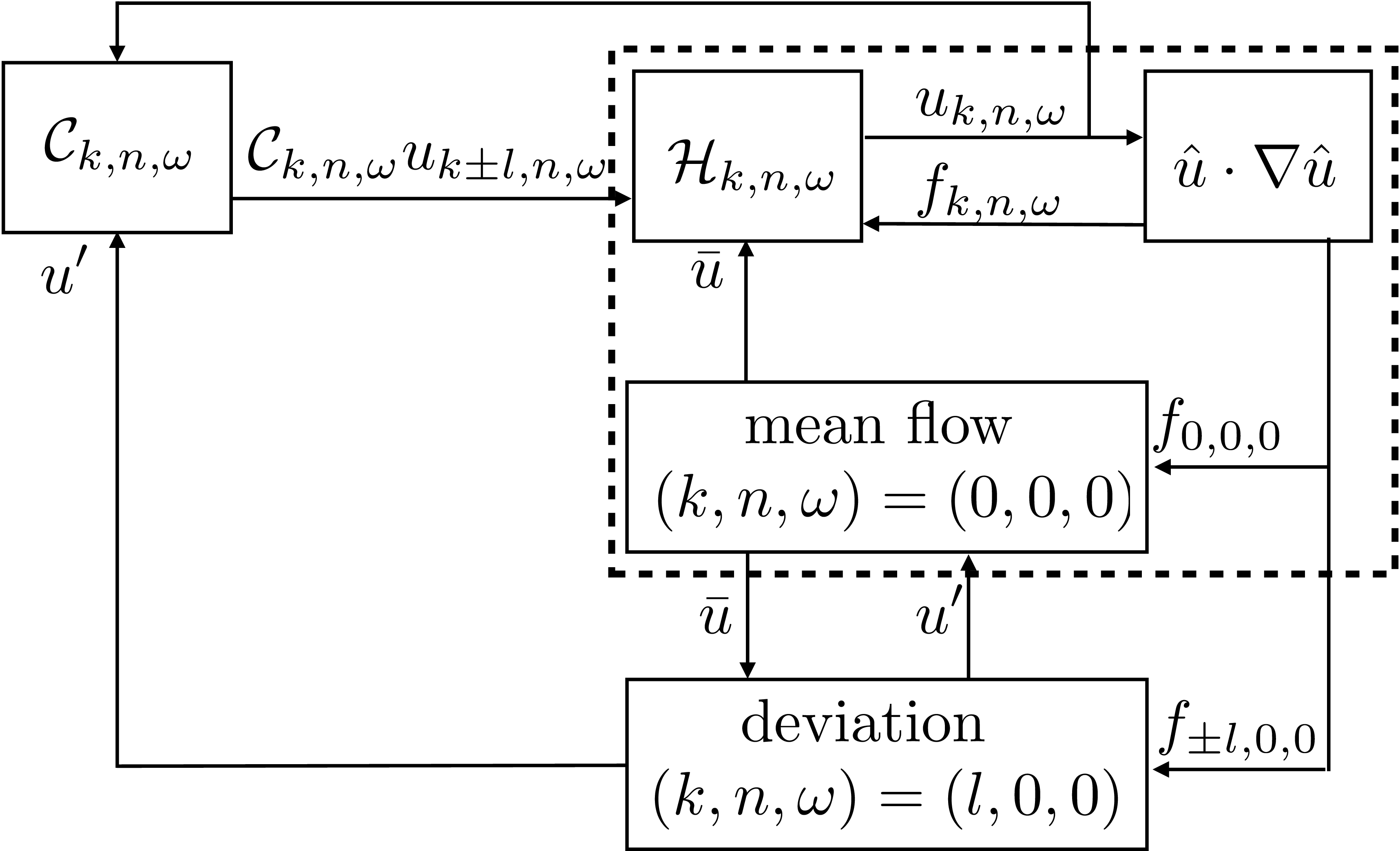}
\caption{\label{fig:15D} Diagram of the new triple-decomposition-based
  resolvent model. The mean velocity profile is sustained in the
  $(k,n,\omega)=(0,0,0)$ equation via the Reynolds stress $f_{0,0,0}$
  and the interactions with deviation velocity. Similarly, the
  deviation velocity is also sustained via the forcing $f_{l,0,0}$ and
  interactions with the mean flow in the $(k,n,\omega)=(l,0,0)$
  equation. The original model of \citet{McKeonSharma2010} is
  represented by the subset within the dashed border. }
\end{center}
\end{figure}

The physical interpretation of the present resolvent formulation is the same to that of the original
formulation \citep{McKeonSharma2010}. Figure~\ref{fig:15D}
presents the new resolvent formulation (\ref{eq.exres}) by means
of a block diagram. The mean velocity profile is sustained in the
$(k,n,\omega)=(0,0,0)$ equation via the Reynolds stress $f_{0,0,0}$
and interactions with deviation velocity. 
Similarly, the deviation velocity is also sustained via the
forcing $f_{l,0,0}$ and interactions with the mean flow in the
mean flow equation corresponding to $(k,n,\omega)=(l,0,0)$. The deviation velocity drives the triadic interactions generated via the operator
$\mathcal{C}_{k,n,\omega}$ and, closing the loop, the mean profile
restricts how the fluctuating velocity responds to the non-linear
forcing via the resolvent operator $\mathcal{H}_{k,n,\omega}$. We note
that it is relatively straightforward to generalize the block diagram in
figure~\ref{fig:15D} if the deviation velocity is composed of multiple
axial wavenumbers, or even frequencies. 

We highlight that this formulation represents the exact route for
modifying a turbulent flow using the mean profile $u_0$. Despite the
non-homogeneity of the flow in our present cases of interest, the operator $\mathcal{H}_{k,n,\omega}$
is identical to the one developed by \citet{McKeonSharma2010} for
\oned\ mean flows and thus it only depends on the axial mean $\bar{u}(r)$;
the deviation velocity does not appear in the resolvent, but it
does appear in the mean flow equation. Also note that the equation of
continuity enforces the condition that the mean profile of wall-normal velocity $v_0$
is zero in all cases.  Hence the modification of the mean profile
$\bar{u}_0$ is sufficient to analyze the dynamics of the flow. We recall that \citet{mckeon2013experimental} used a comparable decomposition in order to assess the effect of a synthetic large-scale motion on the flow dynamics.

Following the analysis of \citet{McKeonSharma2010}, a singular value
decomposition (SVD) of the resolvent operator
\begin{equation}
\mathcal{H}_{k,n,\omega}=\sum_m {\bm{\psi}}_{k,n,\omega,m}
\sigma_{k,n,\omega,m} {\bm{\phi}}^*_{k,n,\omega,m}
\end{equation}
delivers an input-output amplification relation between 
response modes $\bm{\psi}_{k,n,\omega,m}$ and forcing modes
$\bm{\phi}_{k,n,\omega,m}$ through the magnitude of the corresponding
singular value $\sigma_{k,n,\omega,m}$. Here the subscript $m$ is an
index that ranks singular values from largest to smallest. The
non-linear terms $\bm{g}_{k,n,\omega}$ can be decomposed as a sum of
forcing modes to relate the amplification mechanisms to the
velocity fields,
\begin{equation}
\bm{g}_{k,n,\omega}=\sum_m \chi_{k,n,\omega,m} \bm{\phi}_{k,n,\omega,m}
\end{equation}
where the unknown forcing coefficients $\chi_{n,\omega,n}$ represent
the unknown mode interactions and Reynolds stresses. The decomposition
of the fluctuating velocity field is then constructed as a weighted
sum of response modes

\begin{equation}
    \bm{u}(x,r,\theta,t)= \sum_{(k,n,\omega) \neq (l,0,0) }
    \chi_{k,n,\omega,1}\sigma_{k,n,\omega,1}\bm{\psi}_{k,n,\omega,1}\ce^{\ci(kx
      + n\theta-\omega{t})} \, ,
\label{resolde}
\end{equation}
in which the low-rank nature of the resolvent, $\sigma_{k,n,\omega,1}
\gg \sigma_{k,n,\omega,2}$, is exploited
\citep{McKeonSharma2010,sharma2013coherent,luhar2014opposition}.

A numerical method similar to the one developed by
\cite{McKeonSharma2010} is employed for the discretization of the
resolvent operator $ \mathcal{H}_{k,n,\omega}$.  Following the
approach of \citet{meseguer2003linearized}, wall normal derivatives
are computed using Chebychev differentiation matrices, properly modified to avoid the axis
singularity. Notice that instead of projecting the velocity into a
divergence-free basis, here the divergence-free velocity fields are
enforced by adding the continuity equation as an additional column and
row in the discretized resolvent \citep{luhar2014opposition}. The
velocity boundary conditions at the wall are zero Dirichlet. The velocity profile
inputs of the four cases subject to study were shown in
figure~\ref{fig:meanshear}.

\subsection{Fourier analyses and DMD}

It may be observed in (\ref{resolde}) that the energy associated with each Fourier mode ${\bm u}_{k,n,\omega}$ is proportional to its weighting $\chi_{k,n,\omega,1}\sigma_{k,n,\omega,1}$, under the rank-1 assumption. The resolvent analysis yields the amplification properties in $(k,n,\omega)$ but it does not provide information on the amplitude of non-linear forcing $\chi_{k,n,\omega,1}$.

As briefly exposed in \S\ref{sec:intro}, a snapshot-based DMD analysis \citep{Schmid2010,RowleyEtAlJFM2009} is carried out on the DNS data in order to unveil the unknown amplitudes of the non-linear forcing terms $\chi_{k,n,\omega,1}$. DMD obtains the most energetic flow structures, i.e., the set of wavenumber/frequencies $(k,n,\omega)$ corresponding to the maximum product of amplitude forcing and amplification. As shown by \citet{chen2012variants}, the results from a DMD analysis o statistically steady flows such as those considered are equivalent to a discrete Fourier transform of the DNS data once the time-mean is subtracted.  This is confirmed if values of decay/growth of the DMD eigenvalues are close to zero. Hence, the DMD modes obtained are marginally stable, and can be considered Fourier modes. The norms of these modes indicates the energy corresponding to each set of wavenumber/frequencies and can unveil the value of the unknown forcing
coefficients $\chi_{k,n,\omega,1}$ \citep{gomez_iti_2014}.

To avoid additional post-processing of the DNS data in the DMD analysis, we directly employ \twod\ snapshots with Fourier expansions in the azimuthal direction, obtained from the DNS based on (\ref{eq:FourierDecomp}). Given an azimuthal wavenumber $n$, two matrices of snapshots equispaced in time are constructed as

\begin{eqnarray}
\mathcal{U}^1 & = & \left[\begin{array}{cccc} \hat{\bm u}_n(x,r,t_1) & \hat{\bm u}_n(x,r,t_2) & ... & \hat{\bm u}_n(x,r,t_{{N_s}-1})  \end{array} \right] \, , \\
\mathcal{U}^2 & = & \left[\begin{array}{cccc} \hat{\bm u}_n(x,r,t_2) & \hat{\bm u}_n(x,r,t_3) & ... & \hat{\bm u}_n(x,r,t_{N_s})  \end{array} \right] \, .
\end{eqnarray}
The size of these snapshot matrices is $N_rN_x \times{N_s-1}$ with $N_s$ being the number of snapshots employed. DMD consists of the inspection of the properties of the linear operator $\mathcal{A}$ that relates the two snapshot matrices as

\begin{equation}
\mathcal{A} \mathcal{U}^1  = \mathcal{U}^2 \, ;
\end{equation}
the linear operator $\mathcal{A}$ is commonly known as the Koopman operator. In the present case, it can be proved that the eigenvectors of this operator and the Fourier modes of the DNS data are equivalent if the snapshot matrices are not rank deficient and the eigenvalues of $\mathcal{A}$ are isolated. The reader is referred to the work of \citet{chen2012variants} and the review by \citet{mezic2013analysis} for a rigorous derivation of this equivalence. To obtain the eigenvectors of $\mathcal{A}$, we employ the DMD algorithm based on the SVD of the snapshot matrices developed by
\citet{Schmid2010}. This algorithm circumvents any rank-deficiency in the snapshot matrices and it provides a small set of eigenvectors ordered by energy norm. The dataset consist of 1200 DNS
snapshots equispaced over $\mathcal{O}(40)$ wash-out times $L/u_0(R)$.

We anticipate that because of the discretization employed in the DMD analysis, the Fourier modes obtained are \twod\ and can contain multiple axial wavenumbers $k$. A spatial Fourier transform in the axial direction can be carried out in order to identify the dominant axial wavenumber $k$ of a DMD mode.

In the following, we employ the resolvent analysis and DMD to address how the most
amplified and energetic flow structures are manipulated by the
transpiration. This is the first step in establishing a relation
between the changes in flow structures and drag reduction or increase
mechanisms.

\subsection{Amplification and energy}

\begin{figure}
\begin{center}
\includegraphics*[width=1\linewidth]{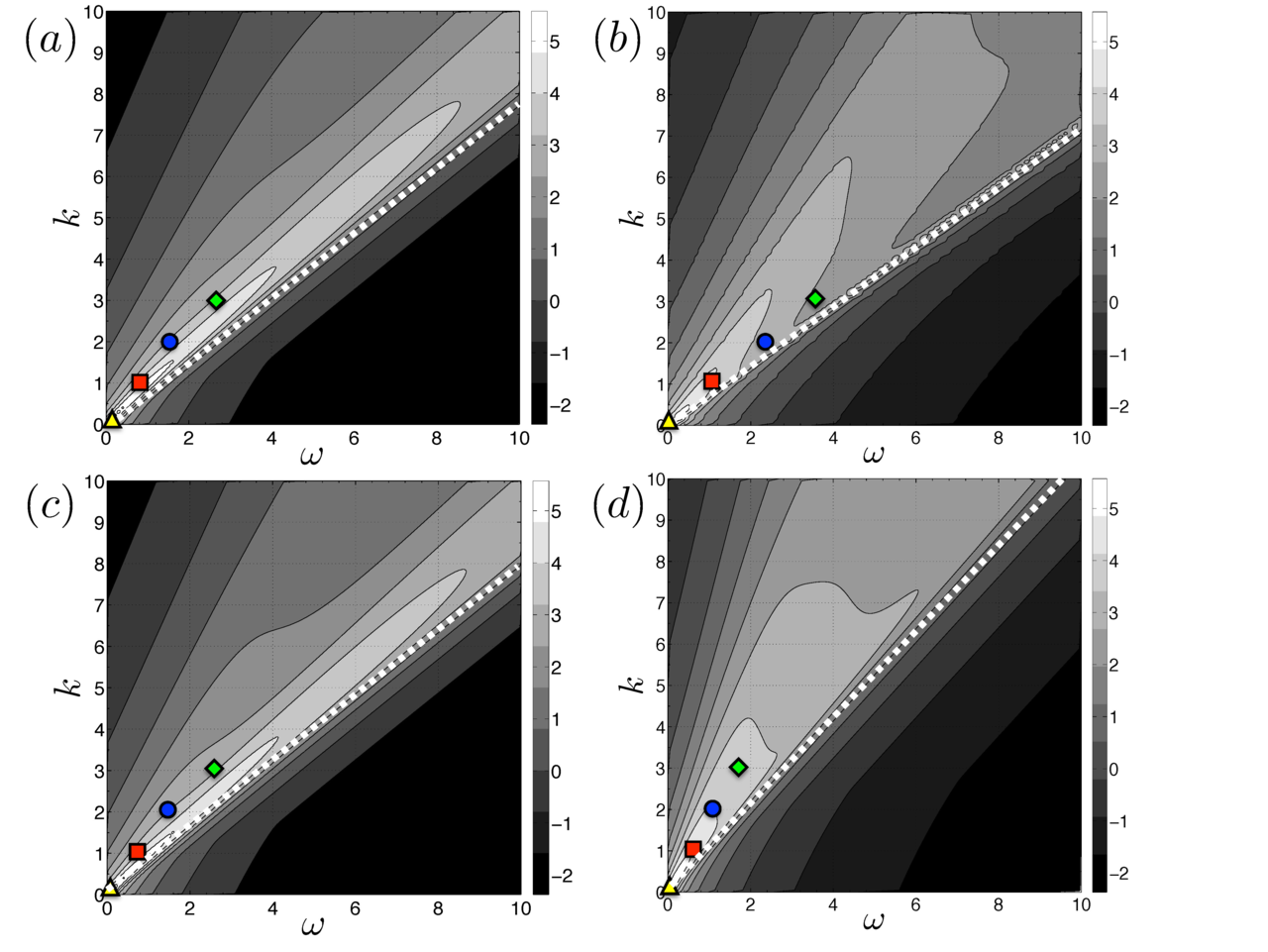}
\caption{\label{fig:sigmas}Distribution of amplification
  $\log_{10}(\sigma_{k,6,\omega,1})$ (\textit{a}) reference: smooth pipe (\textit{b})
  case I: large drag-decrease $(A^+,k_c)=(10,10)$, (\textit{c}) case II: small
  drag-decrease $(A^+,k_c)=(0.7,10)$, (\textit{d}) case III: drag-increase
  $(A^+,k_c)=(2,2)$. Symbols denote the frequency corresponding to the most energetic \twod\ DMD
  modes with $k=0,1,2,3$ dominant wavenumber. Dashed lines indicate the wavespeed corresponding to the centerline velocity $\bar{u}^c(R)$ for each case.}
\end{center}
\end{figure}

\begin{figure}
\begin{center}
\includegraphics*[width=1\linewidth]{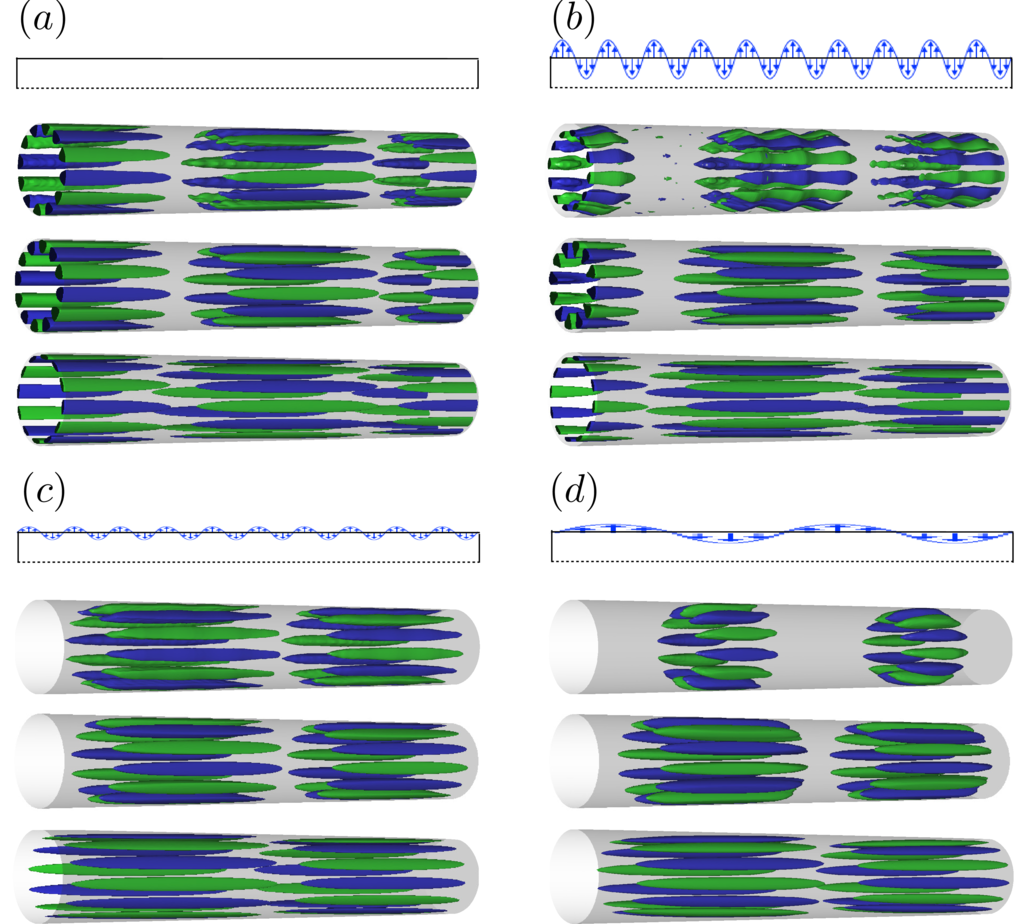}
\caption{\label{fig:DMD} In each of the four blocks, the upper diagram represents each
 transpiration configuration. Below, iso-surfaces of $50 \%$ of the maximum/minimum
  streamwise velocity are represented corresponding to ({\em top}) DMD mode with dominant
  wavenumbers $k=1$ and $n=6$ (red squares in figure~\ref{fig:sigmas})
({\em middle}) DMD modes axially-filtered at $k=1$ ({\em bottom}) resolvent mode associated with $k=1$ and $n=6$ at
  the same frequency. (\textit{a}) reference: smooth pipe (\textit{b})
  case I: large drag-decrease $(A^+,k_c)=(10,10)$, (\textit{c}) case
  II: small drag-decrease $(A^+,k_c)=(0.7,10)$, (\textit{d}) case III:
  drag-increase $(A^+,k_c)=(2,2)$.  }
\end{center}
\end{figure}

We focus on the effect of transpiration on large scale
motions. Hence the broad parameter space $(k,n,\omega)$ is reduced by
taking into consideration findings from the literature. As discussed
in detail by \cite{sharma2013coherent}, VLSMs in pipe flows can be
represented with resolvent modes of lengths scales $(k,n)=(1,6)$ and
with a convective velocity $c=2/3$ of the centerline streamwise
velocity. This representation is based on the work of
\citet{monty2007large} and \cite{bailey2010experimental}, which
experimentally investigated the spanwise length scale associated with
the VLSM and found to be of the order of the outer length scale,
$n=6$. Hence we consider only the azimuthal wavenumber $n=6$.

Reference values of the amplification are provided in contours in
Figure \ref{fig:sigmas}(\textit{a}), which show the distribution of
resolvent amplification $\log_{10}(\sigma_{k,6,\omega,1})$ for the
smooth pipe in a continuum set of $(k,\omega)$ wavenumbers for $n=6$.
We observe a narrow band of high amplification caused by a
critical-layer mechanism \citep{McKeonSharma2010}. This critical-layer
mechanism can be described by examining the resolvent operator. The
diagonal terms of the inverse of the resolvent matrix in
(\ref{discrtH}) read
\begin{equation}
h_{ii} = \left[ \ci (ku_0 - \omega) u + Re^{-1}\nabla^2 u  \right]^{-1} \, ,
\end{equation}
thus, for a given value of the Laplacian, there is a large
amplification if the wavespeed $c=\omega/k$ matches the mean
streamwise velocity, \ie $c=u_0$. That means that flow structures that
travel at the local mean velocity create high
amplification. Furthermore, we note that straight lines that pass
through the origin in figure~\ref{fig:sigmas} correspond to constant
wavespeed values. The wavespeed corresponding to the centerline velocity is indicated by a dashed line.

Turning first to figure~\ref{fig:sigmas}(\textit{a}), the symbols represent the
dominant wavenumber and the frequency corresponding to the four most
energetic DMD modes at $n=6$. The value of the energy corresponding to each the
four modes is similar and omitted.  As explained by \citet{gomez_pof_2014}, a
sparsity is observed in energy as a consequence of the critical layer
mechanism in a finite length periodic domain; only structures with an
integer axial wavenumber, $k_i=1,2,\ldots$ can exist in the flow
 because of the finite length periodic domain with length $L=4\pi
 R$. This fact creates a corresponding sparsity in frequency. For
 each integer axial wavenumber $k_i$ there is a frequency $\omega_i$
 for which the critical layer mechanism occurs, \ie $\omega_i=k_i
  c=k_i u_0(r_c)$, with $r_c$ being the wall-normal location of the
  critical layer. This energy sparsity behaviour is clearly observed
  in the reference case, as shown in
  figure~\ref{fig:sigmas}(\textit{a}), in which the peak frequencies
  are approximately harmonics. Thus they are approximately aligned in
  a constant wavespeed line.

We also observe that, for a given $k$, the frequencies corresponding
to the peaks of energy differ from the most amplified frequency. As observed by \cite{gomez_iti_2014}, this is related to the major role that the non-linear forcing
$\chi_{n,\omega,n}$ maintaining the turbulence plays in the resolvent
decomposition (\ref{resolde}). Nevertheless, we observe that the
frequency corresponding to the most energetic modes can be found in
the proximity of the high amplification band.

Furthermore, the fact there is only one narrow band of amplification
indicates a unique critical layer and that each frequency corresponds
to only one wavenumber. This result is confirmed through the similar
features exhibited by the most energetically relevant flow structures,
arising from DMD of the DNS data, and the resolvent modes associated
with the same frequency and wavenumber.  For instance, figure~\ref{fig:DMD}(\textit{a}) presents a comparison of the DMD and resolvent mode corresponding to $k=1$ (red square in
figure~\ref{fig:sigmas}(\textit{a})).  We observe that both DMD and
resolvent modes present a unique dominant wavenumber $k=1$ with a
similar wall-normal location of its maximum/minimum velocity. The DMD
analysis makes use of \twod\ DNS snapshots based on the Fourier
decomposition in (\ref{eq:FourierDecomp}) and so it permits multiple
axial wavenumbers. Hence, the DMD modes are not constrained to one
unique dominant axial wavenumber. This does not apply to the resolvent
modes, which only admit one wavenumber $k$ by construction of the
model. This uniqueness of a dominant streamwise wavenumber in the most energetic
flow structures of a smooth pipe flow has also been observed in the
work of \citet{gomez_pof_2014}.  Additionally, figure~\ref{fig:DMD}(\textit{a}) presents DMD modes axially-filtered to $k=1$ to ease comparisons with the resolvent modes. We consider it most appropriate to show the axial velocity because it is the energetically dominant
velocity component in these flow structures.  Additionally, the flow
structures corresponding to positive and negative azimuthal wavenumber
$n=\pm 6$ have been summed; a single azimuthal wavenumber necessarily
corresponds to a helical shape.

Figure \ref{fig:sigmas}(\textit{b}) represents the amplification results of
case I, corresponding to large drag-decrease with
$(A^+,k_c)=(10,10)$. The effect of transpiration on the flow dynamics
is significant; two constant-wavespeed rays of high amplification
may be observed. One of the wavespeeds is similar than the one
corresponding the critical layer in the reference case while the
second one is much faster, almost coincident with the centerline velocity. Also the amplification in the area between these two constant wavespeed lines is increased with respect to the reference case. Hence multiple axial wavenumber $k$ could be amplified at each frequency. This
is confirmed in figure~\ref{fig:DMD}(\textit{b}), which shows that the DMD mode
features waviness corresponding to multiple wavenumbers. Although the dominant
axial wavenumber $k=1$ can be
visually identified, other wavenumbers corresponding to interactions of this $k=1$ with the transpiration wavenumber $k_c=10$ are also observed. The waviness in figure~\ref{fig:DMD}(\textit{b}) is steady; it does not travel with the $k=1$ flow structure. This has been confirmed via animations of the flow structures and it was expected as a result of the steady transpiration. 

As mentioned before, the resolvent modes arise from a \oned\ model based
on a $(k,n,\omega)$ Fourier decomposition hence they only contain one axial wavenumber $k$. As such, a single resolvent mode cannot
directly replicate the waviness. However, it provides a
description of the dominant wavenumber flow structure. Nevertheless, the axially-filtered DMD mode in figure~\ref{fig:DMD}(\textit{b}) presents a similar structure to the resolvent mode. Similarly to
the reference case, the wavespeed based on the dominant wavenumber of
the DMD modes are aligned in a constant wavespeed line near the high amplification regions, as seen in \ref{fig:sigmas}(\textit{b}) .

Figure \ref{fig:sigmas}(\textit{c}) shows that the small transpiration
$(A^+,k_c)=(0.7,10)$ does not have a significant influence on the flow
dynamics. The main difference with respect to the reference case is
that the amplification line is slightly broader in this
case. Consistent with this result, the DMD mode in Figure
\ref{fig:DMD}(\textit{b}) shows a tiny waviness corresponding to other wavenumbers. Similarly to the previous cases, the resolvent mode replicates the dominant axial wavenumber
flow structure. 

The amplification and energy results corresponding to the
drag-increase case $(A^+,k_c)=(2,2)$ are shown in
Figure~\ref{fig:sigmas}(\textit{d}). We observe that, in agreement
with the decrease of bulk velocity, the transpiration slows down the
flow dynamics; the wavespeed corresponding to the critical layer is
slower than in previous cases, i.e., a steeper dashed line in
Figure~\ref{fig:sigmas}(\textit{d}).  Additionally, the critical layer
is broad in contrast to the reference case, indicating that multiple axial wavenumbers could be excited. As such, the DMD mode in figure~\ref{fig:DMD}(\textit{d}) shows a dominant axial wavenumber
$k=1$ which interacts with others in order to produce two steady localized areas of fluctuating
velocity. These clusters are located slightly after the two blowing
sections, which corresponding to the high velocity areas in
figure~\ref{fig:means}(\textit{d}). As in previous cases, the resolvent mode
at $k=1$ captures the same dynamics of the filtered DMD mode.

\section{Discussion and conclusions}
\label{sec:end}

The main features of low- and high-amplitude wall transpiration
applied to pipe flow have been investigated by means of direct
numerical simulation at Reynolds number $Re_\tau=314$. Turbulence
statistics have been collected during parameter sweeps of different
transpiration amplitudes and wavenumbers. The effect of transpiration
is assessed in terms of changes in the bulk axial flow.

We have shown that for low amplitude transpiration the mean streamwise
velocity profile follows a velocity defect law, so the outer flow is
unaffected by transpiration. This indicates that the flow still obeys
Townsend's similarity hypothesis \citep{Townsend1976}, as it is
usually observed in rough walls.  Hence, low amplitude transpiration
has a similar effect as roughness or corrugation in a pipe. On the
other hand, high amplitude transpiration has a dramatic effect on the
outer layer of the velocity profile and the overlap region is
substituted by a large increase in streamwise velocity.

We have observed that a transpiration configuration with a small
transpiration wavenumber leads to long regions of suction in which the
streamwise mean velocity is significantly reduced and the fluctuating
velocity can be suppressed. In contrast, a large transpiration
wavenumber can speed up the outer layer of the streamwise mean profile
with respect to the uncontrolled pipe flow, even at small
amplitudes.

These trends in amplitude and wavenumber have permitted the identification of
transpiration configurations that lead to a significant drag increase
or decrease. For instance, we have shown that a wall transpiration
that combines a large amplitude with a large wavenumber creates a
large increase in flow rate.

A comparison with the channel flow data of \citet{quadrio2007effect}
revealed that application of low amplitude transpiration to the pipe
flow leads to similar quantitative results.

The obtained turbulence statistics showed that the changes in Reynolds
stress induced by the transpiration are not sufficient to explain the
overall change in the mean profile. An analysis of the streamwise
momentum equation revealed three different physical mechanisms that
act in the flow: modification of Reynolds shear stress, steady
streaming and generation of non-zero mean streamwise gradients.
Additionally, a triple decomposition of the velocity based on a mean
profile, a \twod\ deviation from the mean profile and a fluctuating
velocity, has been employed to examine the streamwise momentum
equation. This decomposition showed that the steady streaming term can
be interpreted as a coherent Reynolds stress generated by the
deviation velocity. The contribution of this coherent Reynolds stress
to the momentum balance is important close to the wall and it affects
the viscous sublayer, which is no longer linear under the influence of
transpiration.

The behaviour of these terms has been examined by selected
transpiration cases of practical interest in terms of drag
modification. For all cases considered, the steady streaming terms
opposes to the flow rate while the change in Reynolds shear stress are
always positive. This concurs with the numerical simulations of
\citet{quadrio2007effect,hoepffner2009pumping} and the perturbation
analysis of \citet{woodcock2012induced}. Additionally, we have observed
that the contribution of non-zero mean streamwise gradients is
significant. This contribution can be negative for small transpiration
wavenumbers; the turbulent fluctuations are suppressed in the large
areas of suction, favoring non-homogeneity effects in the axial
direction.

A description of the change in the flow dynamics induced by the
transpiration has been obtained via the resolvent analysis methodology
introduced by \citet{McKeonSharma2010}. This framework has been
extended to deal with pipe flows with an axially-invariant
cross-section but with mean spatial periodicity induced by changes in
boundary conditions. The extension involves a triple decomposition
based on mean, deviation and fluctuating velocities. This new
formulation opens up a new avenue for modifying turbulence using only
the mean profile and it could be applied to investigate the flow
dynamics of pulsatile flows or changes induced by dynamic roughness.

In the present investigation, this input--output analysis showed that
the critical-layer mechanism dominates the behaviour of the
fluctuating velocity in pipe flow under transpiration. However,
axially periodic transpiration actuation acts to delocalize the
critical layer by distorting the mean flow, so that multiple
wavenumbers can be excited. This produces waviness of the flow
structures.

The resolvent results in this case are useful as a tool to interpret
the dynamics but are less directly useful to predict the effects of
transpiration, since transpiration feeds directly into altering
the mean flow, which itself is required as an input to the resolvent
analysis. This limitation partly arises owing to the use of steady
actuation, since low-amplitude time-varying actuation directly forces
inputs to the resolvent rather than altering its structure (see
figure~\ref{fig:15D})

The critical layer mechanism concentrates the response to actuation in
the wall-normal location of the critical layer associated with the
wavespeed calculated from the frequency and wavenumber of the
actuation.  This leads us to believe that, within this framework, dynamic actuation may be
more useful for directly targeting specific modes in localised regions
of the flow.

DMD of the DNS data confirmed that the transpiration mainly provides a
waviness to the leading DMD modes. This corrugation of the flow
structures is steady and corresponds to interactions of the critical-layer induced wavenumber with the transpiration wavenumber. As such, this waviness is responsible for generating
steady streaming and non-zero mean streamwise gradients, which it turns modifies the streamwise momentum balance, hence enhancing or decreasing the drag. 

Finally, a performance analysis indicated that all the
transpiration configurations considered are energetically inefficient. In the
most favorable case, the benefit obtained by the drag reduction
induced by the transpiration marginally exceeds the cost of applying
transpiration. Nevertheless, this is an open loop active flow control
system. A passive roughness-based flow control system able to mimic
the effect of transpiration would be of high practical interest.
Hence, we remark that experience gained through this investigation
serves to extend this methodology towards manipulation of flow
structures at higher Reynolds numbers; this is the subject of an
ongoing investigation.

\section*{Acknowledgments}
The authors acknowledge financial support from the Australian Research
Council through the ARC Discovery Project DP130103103, from
Australia's National Computational Infrastructure via Merit Allocation
Scheme Grant d77, and from the U.S. Office of Naval Research, grant \#N000141310739 (BJM).

\section*{Appendix A. Triadic interaction induced by the deviation velocity}

Without loss of generality, we write the deviation velocity as a
single Fourier mode with $l$ axial wavenumber. Then the triple
decomposition reads:
\begin{equation}
\hat{\bm u}(x,r,\theta,t)=\underbrace{\bar{\bm u}(r)}_{\mbox{A}} +
\underbrace{{\bm u}_l^\prime(r) e^{\mathrm{i}lx}}_{\mbox{B}} +
\overbrace{\sum_{(k,n,\omega) \neq (l,0,0)} {\bm u}_{k,n,\omega}(r)
  e^{\mathrm{i}(kx+n\theta-\omega t)}}^{\mbox{C}} + \CC
\end{equation}
As an example, we substitute this decomposition in the nonlinear terms
$\hat{u} \partial_x \hat{u}$
\begin{equation}
\hat{u} \partial_x \hat{u}=A \partial_x A +A \partial_x B + A
\partial_x C + B \partial_x A + B \partial_x B + B \partial_x C + C
\partial_x A + C \partial_x B + C \partial_x C \, ,
\end{equation}
since $\partial_x A = 0$ we obtain 
\begin{equation}
\hat{u} \partial_x \hat{u}= A \partial_x B + A \partial_x C + B
\partial_x B + B \partial_x C + C \partial_x B + C \partial_x C \, .
\end{equation}
Next, we study which terms are orthogonal to the complex exponential functions corresponding to $(k,n,\omega) \neq (0,0,0)$ and $(k,n,\omega) \neq (l,0,0)$. Thus

\begin{equation}
\hat{u} \partial_x \hat{u}= \cancel{A \partial_x B} + \underbrace{A
  \partial_x C}_{\mbox{$k \bar{u} u_{k,n,\omega}$}} + \cancel{B
  \partial_x B} + \overbrace{B \partial_x C}^{\mbox{$(k\pm
    l){u^\prime_l} u_{k\pm l,n,\omega}$}} + \underbrace{C \partial_x
  B}_{\mbox{$\pm l {u^\prime_l} u_{k \pm l,n,\omega}$}} + \overbrace{C
  \partial_x C}^{\mbox{$f_{k,n,\omega}$}} \, .
\end{equation} 
As a consequence of the triadic interaction between the spatial
Fourier mode of the deviation velocity and the fluctuating velocity,
there is a coupling of the fluctuating velocity at $(k,n,\omega)$ with
that at $(k \pm l,n,\omega)$. This interaction generates the new terms
$B \partial_x C$ and $C \partial_x B$. These two terms are represented
by the coupling operator $\mathcal{C}_{k,n,\omega}$ in the resolvent
formulation (\ref{coupling}). Note that the term $A \partial_x B$
is included in the deviation equation $(k,n,\omega) = (l,0,0)$ while
the term $B \partial_x B$ contributes to the mean flow equation
$(k,n,\omega) = (0,0,0)$ as coherent Reynolds stress.

\bibliographystyle{jfm}
\bibliography{jfm}

\end{document}